\documentclass[journal]{IEEEtran}

\usepackage{amsfonts}
\usepackage{ifpdf}
\usepackage{xcolor}
\usepackage{soul}
\usepackage{cite}
\usepackage{balance}
\usepackage{hyperref}
\hypersetup{colorlinks,linkcolor={black},citecolor={black},urlcolor={black}}

\ifCLASSINFOpdf
   \usepackage[pdftex]{graphicx}
\else
   \usepackage[dvips]{graphicx}
\fi

\usepackage{color}
\usepackage{amsmath}
\usepackage{amsthm}
\usepackage{bm}

\DeclareMathOperator*{\argmax}{arg\,max}

\usepackage[linesnumbered,ruled,vlined]{algorithm2e}
\let\oldnl\nl
\newcommand{\nonl}{\renewcommand{\nl}{\let\nl\oldnl}}
\SetKwInput{KwInput}{Input}
\SetKwInput{KwOutput}{Output}

\usepackage{algorithmicx, algpseudocode}
\usepackage{array}
\usepackage{fixltx2e}
\usepackage{stfloats}
\usepackage{url}
\begin{document}

\title{Robust and Feasible QoS-Aware mmWave Massive MIMO Hybrid Beamforming}

\author{Mohsen~Tajallifar,~Ahmad~R.~Sharafat,~\IEEEmembership{Life~Senior~Member,~IEEE},~and Halim~Yanikomeroglu,~\IEEEmembership{Fellow,~IEEE}
\thanks{M. Tajallifar and A. R. Sharafat are with the Faculty of ECE, Tarbiat Modares University, Tehran, Iran. H. Yanikomeroglu is with the Department of SCE, Carleton University, Ottawa, Canada. 
	
	 Corresponding author is A. R. Sharafat (e-mail: sharafat@ieee.org).}}

\maketitle

\begin{abstract}
Hybrid beamforming (HB) with quality-of-service (QoS) provisioning per stream in millimeter waves is indispensable in 5G/6G networks. HB includes baseband and radio frequency (RF) beamforming, and requires error-free channel state information (CSI), which is erroneous in practice. So there is a need for efficient, feasible, robust, and QoS-aware HB. To achieve this, we mitigate CSI uncertainty via baseband beamforming, and we steer the RF beamformer by using the estimates of the channel's eigenvectors. In doing so, we consider the effective channel's uncertainty region instead of the uncertainty region of the channel itself, as the former is smaller than the latter, requiring less transmit power to satisfy the QoS constraint. We also detect and eliminate the infeasible data streams. Our iterative scheme (which is based on the cutting-set method) for baseband beamforming satisfies the mean-squared error (MSE) constraint per stream, where a limited number of constraints are considered instead of infinitely many constraints. In our low-complexity scheme, we derive a simple sufficient condition to check the feasibility of each stream, we diagonalize the effective channel at the baseband precoder, and we use minimum MSE combining at the baseband combiner. Extensive simulations validate our formulations and theoretical derivations.
\end{abstract}

\IEEEpeerreviewmaketitle
\section{Introduction}\label{introduction_sec}

Millimeter waves (mmWaves) resolve {bandwidth bottlenecks in mobile networks} but suffer from substantial path loss, which can be {compensated for by the} array gain of massive multi-input multi-output (MIMO) systems. A conceptually simple scheme {for implementing massive MIMO systems} is to assign a dedicated radio frequency (RF) chain to each antenna element, {but this requires additional} hardware, consumes excessive power, and is costly. 

Hybrid beamforming (HB) is an enabling technology that leverages mmWaves and massive MIMO at a reasonable cost. HB is performed in the baseband and RF domains. In the baseband domain, beamforming is performed with full control over the signal's amplitude and phase. In the RF domain, analog beamforming is performed by utilizing phase shifters only, where phase shifts are applied to the signal, {and each RF chain is linked to multiple antenna elements.}

HB has been a topic of extensive research in recent years \cite{el2014spatially, yu2016alternating, sohrabi2016hybrid, ioushua2019family, ni2015hybrid, wu2018hybrid, wang2022joint, ardah2020hybrid, tsinos2017onthe, zang2019optimal, ni2017energy, shen2017joint}. In \cite{el2014spatially, yu2016alternating, sohrabi2016hybrid, ioushua2019family, ni2015hybrid, wu2018hybrid, wang2022joint, ardah2020hybrid, tsinos2017onthe}, HB's objective is to optimize a global measure, {such as the sum-rate} in all data streams, {or the sum mean-squared error (MSE) of all data streams,} or energy efficiency. In one approach {for implementing HB}, the hybrid precoder and the hybrid combiner are approximations of their fully digital counterparts \cite{el2014spatially,yu2016alternating}{; where approximations were obtained via the orthogonal matching pursuit algorithm in} \cite{el2014spatially} {and manifold optimization in} \cite{yu2016alternating}. Another approach {decoupled} the design of the baseband digital precoder from that of {the RF analog precoder. The latter was obtained by maximizing spectral efficiency, and the former was found by orthogonalizing the effective channel} \cite{sohrabi2016hybrid}. 

HB is also performed in {the} quality of service (QoS)-aware systems, where the precoder and combiner minimize the transmit power while satisfying the QoS constraint of each data stream, {for instance the} signal-to-interference-plus-noise ratio (SINR) \cite{zang2019optimal} or {the data rate} \cite{ni2017energy, shen2017joint}.

The aforementioned HB schemes assume perfect channel state information (CSI), which is unrealistic {in practice} due to channel estimation errors, outdated CSI, pilot contamination, and quantization. CSI error is more severe in mmWave massive MIMO systems {on account of} the large size of channel matrices and {short} coherence time, which {has been} shown to be the case in unmanned aerial vehicles, where high mobility, airflow disturbances, and engine vibrations make CSI acquisition even more difficult \cite{xiao2021survey}. {Assuming CSI to be error-free when it is actually erroneous,} may cause considerable performance degradation \cite{zhang2008statistically, pascual2005robust}. To mitigate CSI uncertainty, robust beamforming can be realized either by utilizing the stochastic or the worst-case methods \cite{boyd2004convex}.

In the stochastic approach \cite{zhang2008statistically}, CSI error is assumed to be random with a known probability distribution, and performance is statistically optimized. Channel estimation error is assumed to be additive with Gaussian distribution, and different performance measures have been considered (e.g., the average received SNR \cite{luo2018robust, luo2018robust_1, sun2019robust}, the average sum-rate \cite{thomas2020rate}, the average mutual information \cite{jiang2019mmwave, luo2020robust_2}, and the average sum MSE \cite{jagyasi2016low, jagyasi2017low_1, mai2018hybrid, jagyasi2018band}.) In \cite{kolawole2018rate, sun2019robust, yang2020bayesian}, the parameter of interest {was} the error in the estimated angle of arrival or angle of departure. Moreover, the performance measure {was the instantaneous SINR in} \cite{xu2017outage} (or the {interference in} \cite{ yuan2020hybrid}), which needs to be higher in \cite{xu2017outage} (or lower in \cite{ yuan2020hybrid}) than a given threshold with a predetermined probability. In the stochastic approach, only statistical values of {the performance measure} are considered, which means that QoS constraints may be violated in some instances. {For this reason,} the stochastic approach is not appropriate for QoS-aware applications.

In the worst-case approach, {errors in CSI are assumed for} a predetermined uncertainty region \cite{pascual2005robust}, and performance is optimized for the worst-case CSI errors in that region. For optimizing a global measure, the worst-case robust HB maximizes a lower bound or minimizes an upper bound of the measure. For example, the maximum of sum MSE {was} minimized (min-max optimization) in \cite{morsali2019robust} and the minimum secrecy rate {was} maximized in \cite{cai2020secure}.

In the worst-case approach, QoS constraints are satisfied for all CSI realizations in the uncertainty region, resulting in infinitely many QoS constraints. This may cause the problem to be infeasible for some CSI realizations \cite{vucic2009robust}. In \cite{li2017energy}, the transmit power {was} minimized under worst-case SINR constraint for each user, but feasibility {was} not analyzed.

\begin{figure*}
	\centering
	\includegraphics[width=15.3 cm]{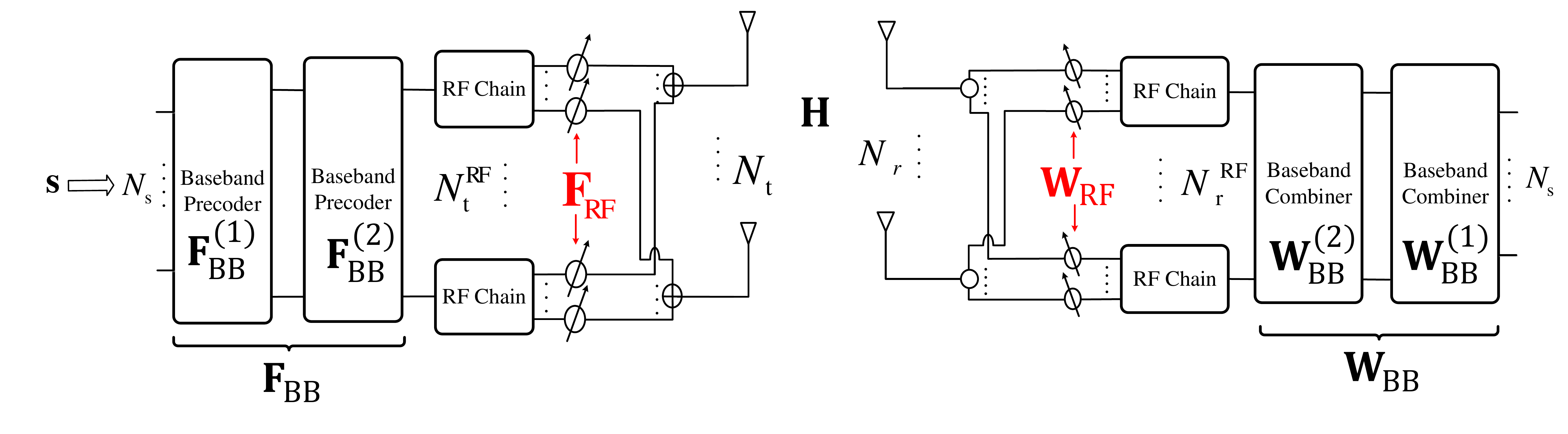}
	\vspace{-6 mm}
	\caption{{Block diagram of a two-stage hybrid beamformer with fully connected arrays.}}
	\label{hybrid_fig}
	\vspace{-6 mm}
\end{figure*}

In this paper, we consider QoS-aware HB in single- and multi-user systems in the presence of CSI uncertainty, and {we} minimize transmit power subject to individual MSE constraints for different data streams. In doing so, we adopt the worst-case approach{; that is, we} guarantee that the MSE of each data stream is below a predetermined threshold for all channel realizations in a given uncertainty region. In what follows, we explain our approach in more detail.

Although eigen-beamforming in fully digital beamformers is efficient for min-max optimization \cite{wang2010robust, wang2011joint}, it cannot be applied to hybrid beamformers because each entry in the matrix of HB's phase shifts in the RF domain has a unit amplitude. In order to {apply eigen-beamforming} to hybrid beamformers, we utilize the two-stage cascade structure in \cite{tajalli2021qos}, and {we} obtain the RF beamformer by approximating the right and left eigenvectors of {the} estimated channel matrix. {The two-stage structure for HB in} \cite{tajalli2021qos}, {shown in Fig.} \ref{hybrid_fig}, {assumes perfect CSI, which is not attainable in practice. By contrast, in this paper, we assume CSI to be uncertain, and we develop a robust HB scheme for the two-stage structure to satisfy the QoS constraint with minimal transmit power, but with a completely different baseband beamforming than} \cite{tajalli2021qos} {that does not suffer from infeasibility and has finite constraints.} 

{In our robust baseband beamformer, we obtain the precoder's matrix of stage 1 in the two-stage cascade structure. To do so, we map the channel's uncertainty region to that of the effective channel; that is we guarantee that the effective channel is in its uncertainty region when the channel itself is in its own uncertainty region.}

The baseband beamformer is designed to deal with infeasibility and infinitely many constraints. When feasible solutions cannot be obtained, we find and eliminate infeasible constraints. {To do so, we add the new variable $\alpha_k$ to each data stream $k$, penalize the objective function with $\sum_{k} \alpha_k$, and obtain the values of $\alpha_k$ by solving the newly formulated optimization problem.} Only those streams with $\alpha_k=0$ are feasible. In this way, the problem becomes feasible \cite{Chinneck2008feasibility}. We also address the infinitely many constraints by the cutting-set method \cite{mutapcic2009cutting}, wherein the worst-case channel for each MSE constraint is obtained via semi-definite programming (SDP).

We also propose a low-complexity method to obtain baseband precoder and combiner matrices in closed form, {which involves diagonalizing} the effective channel at the baseband precoder and {utilizing} minimum MSE (MMSE) combining at the combiner. Next, we obtain a sufficient condition for {the} feasibility of each data stream, compute {the} worst-case MSE for each feasible data stream, and allocate transmit power {for satisfying the} MSE constraint.
\vspace{-2 mm}
\subsection{Contribution and Technical Novelty}

{In this paper, we} develop a robust, feasible and QoS-aware HB scheme in the presence of CSI uncertainty. Our main contributions and achievements are as follows:
\begin{itemize}
\item We map the channel's uncertainty region to the effective channel's uncertainty region and show that the latter is much smaller than the former with the same outage probability.
\item We obtain a feasible subset of data streams by finding and eliminating those data streams that cause infeasibility.
\item We use the cutting-set method to address CSI uncertainty, and {we} find robust HB matrices by alternating between MSE maximization and transmit power minimization. We formulate MSE maximization as a semi-definite programming (SDP) porblem and transmit power minimization as a quadratically constrained quadratic programming (QCQP) problem.
\item We develop a low-complexity HB scheme in which we derive a simple sufficient feasibility condition for each MSE constraint, {while eliminating} those data streams that cause infeasibility. In this scheme, we diagonalize the effective channel in the precoder, utilize MMSE combining in the combiner, and derive a closed-form solution for transmit power allocation.
\item Simulations show that our robust schemes consume less transmit power than the fully digital robust beamforming schemes.
\end{itemize}
\vspace{-4 mm}
\subsection{Organization}
The rest of this paper is organized as follows. The system and channel models are {described} in Section \ref{system_sec}, and {the} QoS-aware HB is formulated in Section \ref{prob_sec}. Our scheme for RF beamforming is in Section \ref{analog_section}, and the uncertainty region for the effective channel is discussed in Section \ref{set_section}. Our iterative and low-complexity schemes for baseband precoding and combining are {described} in Sections \ref{iterative_section} and \ref{low_section}, respectively. Our robust schemes for multi-user (MU)-MIMO HB are {introduced} in Section \ref{mu_section}, followed by simulations {in Section} \ref{simulation_section} {and concluding remarks in Section} \ref{conclusion_section}.
\vspace{-2 mm}
\subsection{Notations}
The following notations are used in this paper: $\bf A$, $\bf a$, and $a$ denote a matrix, a vector, and a scalar, respectively; {$\mathbb E_n$ is the statistical expectation over random variable $n$}; and $\mathbb C^{n \times m}$ is {an} $n\times m$ matrix with complex elements. The operators {$\| \bf A\|_{\rm F}$, $(\bf A)^{\rm *}$, $(\bf A)^{\rm T}$, $(\bf A)^{\rm H}$, $(\bf A)^{-1}$, and $\rm Tr(\bf A)$ are the Frobenius norm, conjugate, transpose, Hermitian, inverse, and trace of  $\bf A$, respectively; $|a|$, $\mathcal{R}\{a\}$, $\mathcal{I}\{a\}$ are the absolute, real, and imaginary values of $a$; $[{\bf A}]_{1:n}$, $[{\bf A}]_{n,m}$, and $\boldsymbol \lambda (\bf A)$ are the first $n$ columns, $(n,m)$ entry, and the vector of eigenvalues of $\bf A$, respectively; ${\bf 1}_N$ is a vector of $N$ ones; $\rm diag\{\bf a\}$ is a diagonal matrix with diagonal elements $\bf a$; and ${\rm diag}({\bf A}_1,\ldots,{\bf A}_n)$ is a block diagonal matrix whose diagonal blocks are ${\bf A}_1,\ldots,{\bf A}_n$.}
\vspace{-20 pt}
\section{System and Channel Models}\label{system_sec}

\subsection{System Model}\label{single-user}

{We consider a single-user system as shown in Fig.} \ref{hybrid_fig}, where the transmitter and receiver include $N_{\rm{t}}$ and $N_{\rm{r}}$ antennas, respectively. A vector ${\bf{s}}\in \mathbb{C}^{N_{\rm{s}} \times 1}$ of $N_{\rm{s}}$ independent data streams with covariance matrix $\mathbb{E}\{{\bf{s}}{{\bf{s}}^{\rm{H}}}\} = {{\bf{I}}_{{N_{\bf{s}}}}}$ is transmitted to the receiver by feeding it into the baseband precoder ${\bf{F}}_{\rm{BB}}\in \mathbb{C}^{N^{\rm{RF}}_{\rm{t}}\times N_{\rm{s}}}$, and passing the results through $N^{\rm{RF}}_{\rm{t}}$ {RF chains. The RF signals are fed into the RF precoder} comprising a set of phase shifters ${\bf{F}}_{\rm{RF}}\in \mathbb{C}^{N_{\rm{t}}\times N^{\rm{RF}}_{\rm{t}}}$, {and the results are sent} to the transmitting antennas. We assume $N^{\rm{RF}}_{\rm{t}}<N_{\rm{t}}$ and $N_{\rm{s}}\le N^{\rm{RF}}_{\rm{t}}$ to ensure signal recovery at the receiver. Similar assumptions are made for combiner at the receiver.

Without loss of generality and for mathematical ease, we assume narrow-band fading channels, {but extending this} to wide-band channels is straightforward. The signal at the receiver's antenna array is
\begin{equation}\label{x}
{\bf{x}} = {{\bf{H}}}{ \bf{F}}_{\rm{RF}}{\bf{F}}_{\rm{BB}}{\bf{s}} + {{\bf{n}}},
\end{equation}
where ${\bf{H}}\in \mathbb{C}^{N_{\rm{r}}\times N_{\rm{t}}}$ is the channel matrix, and ${{\bf{n}}}\in \mathbb{C}^{N_{\rm{r}} \times 1}$ is the noise vector with covariance matrix $\mathbb{E}\{{\bf{n}}{{\bf{n}}^{\rm H}}\} =\sigma^{2}_{\rm{n}} {{\bf{I}}_{{N_{\bf{r}}}}}$. {The extended Saleh-Valenzuela channel model in} \cite{tajalli2021qos} {is obtained by estimating its parameters at the receiver. When the channel is reciprocal, the estimated parameters are sent back to the transmitter.}

The signal $\bf s$ is estimated at the receiver by feeding $\bf x$ in \eqref{x} into the RF combiner comprising a set of phase shifters ${\bf{W}}_{\rm{RF}}\in \mathbb{C}^{N_{\rm{r}}\times N^{\rm{RF}}_{\rm{r}}}$, passing the results through $N^{\rm{RF}}_{\rm{r}}$ RF chains, and feeding the outcome into the baseband combiner ${\bf{W}}_{\rm{BB}}\in \mathbb{C}^{N^{\rm{RF}}_{\rm{r}}\times N_{\rm{s}}}$, resulting in
\begin{equation}\label{s_eq}
\hat{\bf{s}} = {\bf{W}}_{\rm{BB}}^{\rm H}{\bf{W}}_{\rm{RF}}^{\rm H}{{\bf{H}}}{ \bf{F}}_{\rm{RF}}{\bf{F}}_{\rm{BB}}{\bf{s}} + {\bf{W}}_{\rm{BB}}^{\rm H}{\bf{W}}_{\rm{RF}}^{\rm H}{{\bf{n}}}.
\end{equation}

{We consider both the fully connected structure (FCS) and the partially connected structure (PCS) for connecting the RF chains to antenna elements. In the FCS, each RF chain is connected to all antenna elements through $N_{\rm t}$ phase shifters. Hence, the entry $(m,l)$ in ${\bf{F}}_{\rm{RF}}$ and ${\bf{W}}_{\rm{RF}}$ is in the form $e^{j\varphi_{ml}}$, where ${\varphi_{ml}}$ is the phase shift applied to the RF chain $l$ connected to antenna element $m$. In the PCS, each RF chain is connected to a subset of antennas (i.e., a sub-array). The analog precoder in the PCS is ${\bf{F}}_{\rm{RF}} = {\rm diag}\left({\bf{f}}_{\rm{RF},1},\ldots, {\bf{f}}_{\rm{RF}, N_{\rm t}^{\rm RF}}\right)$, where ${\bf{f}}_{{\rm{RF}},l}\in \mathbb{C}^{M \times 1}$ is a vector of $M$ phase shifts applied to the antenna elements in sub-array $m$. The sets of matrices for RF precoders and combiners in the FCS are $\mathcal{F}_{\rm FC}$ and $\mathcal{W}_{\rm FC}$, respectively, and in the PCS are $\mathcal{F}_{\rm PC}$ and $\mathcal{W}_{\rm PC}$, respectively.}\vspace{-2 mm}

\subsection{CSI Uncertainty}\label{CSI model}
In practice, only the estimate of $\bf H$ is available, with {the} additive estimation error, i.e.,\vspace{-2 mm} 
\begin{equation}\label{channel_error}
{\bf H}= {\bf \widehat H}+\boldsymbol\Delta,
\end{equation}
where ${\bf \widehat H}$ is the estimated channel matrix, and $\boldsymbol\Delta$ is the estimation error matrix. The error is norm-bounded, and ${\boldsymbol\Delta}$ is in a hypersphere with the radius of $\varepsilon$ (i.e., the uncertainty region is ${\mathcal E}= \{\boldsymbol\Delta| \| \boldsymbol\Delta \|_{\rm F}\le \varepsilon \}$). In practice, $\varepsilon$ is determined so that ${\rm Pr}\{\boldsymbol\Delta \in  {\mathcal E}\} =  {\rm Pr}\{\|\boldsymbol\Delta \|_{\rm F}\le \varepsilon \} =  P_{\rm in}$. {In other words, the} error is in the uncertainty region with probability $P_{\rm in}$. When $[\boldsymbol\Delta]_{i,j} \sim \mathcal{CN}(0,\sigma^2_{\rm e})$ and the entries of $\boldsymbol\Delta$ are independent and identically {distributed} (i.i.d.), $\| \boldsymbol\Delta \|_{\rm F}^2$ has Erlang distribution with shape $ N_{\rm t} N_{\rm r}$ and rate $\sigma^{-2}_{\rm e}$ \cite{papoulis1989probability}. So, we have $\varepsilon^2 = \phi^{-1}_1(P_{\rm in})$, where $\phi_1(\cdot)$ is the cumulative distribution function (CDF) of $\| \boldsymbol\Delta \|_{\rm F}^2$. Note that $\| \boldsymbol\Delta \|_{\rm F}^2$ is the sum of $ N_{\rm t} N_{\rm r}$  i.i.d. random variables. In massive MIMO systems, $N_{\rm t} N_{\rm r}$ is large, and from the central limit theorem, the probability distribution of $\| \boldsymbol\Delta \|_{\rm F}^2$ is approximated by Gaussian distribution even when the distribution of $\boldsymbol\Delta$ is not Gaussian.
\vspace{-8 pt}
\section{Problem Formulation}\label{prob_sec}
We wish to minimize the transmit power while the MSE constraint as a QoS measure is satisfied for all instances of channel error in the uncertainty region. Note that the obtained value of MSE is valid for any signaling type, e.g., Gaussian, quadrature phase shift keying, and quadrature amplitude modulation. The value of MSE can be converted to other performance measures, e.g., SINR, sum rate, and bit error rate, for a given signaling as described in \cite{palomar2007mimo}. The transmit power is
\begin{equation}\label{power_eq}
P_{\rm T} = {\rm Tr}\left(  {\bf{F}}_{\rm{BB}}^{\rm H}{\bf{F}}_{\rm{RF}}^{\rm H}{\bf{F}}_{\rm{RF}}{\bf{F}}_{\rm{BB}} \right).
\end{equation}
The MSE matrix is ${\bf E}=\mathbb{E}_{\bf s,n}\{(\hat{\bf s} - {\bf s}) (\hat{\bf s} - {\bf s})^{\rm H}\}$, where $[{\bf E}]_{k,k}$ is {the} MSE in the estimate of data stream $k$. {When $\hat{\bf s}$ is as in} \eqref{s_eq}, we have
\begin{equation}\label{mse_eq}
{\bf E}=\left({\bf{W}}^{\rm H}{{\bf{H}}}{ \bf{F}}-{\bf I}_{N_s}\right) 
\left({\bf{W}}^{\rm H}{{\bf{H}}}{ \bf{F}}-{\bf I}_{N_s}\right)^{\rm H} + \sigma^2_{\rm n} {\bf{W}}^{\rm H}{\bf{W}},
\end{equation}
where ${\bf{W}}={\bf{W}}_{\rm{BB}}{\bf{W}}_{\rm{RF}}$ and ${\bf{F}}={\bf{F}}_{\rm{BB}}{\bf{F}}_{\rm{RF}}$. Plugging \eqref{channel_error} into \eqref{mse_eq} results in
\begin{multline}\label{mse_wc}
{\bf E}({\bf{F}},{\bf{W}};{\boldsymbol \Delta})=\left({\bf{W}}^{\rm H}{({\bf \widehat H}+\boldsymbol\Delta)}{ \bf{F}}-{\bf I}_{N_s}\right) \times \\
\left({\bf{W}}^{\rm H}{({\bf \widehat H}+\boldsymbol\Delta)}{ \bf{F}}-{\bf I}_{N_s}\right)^{\rm H} + \sigma^2_{\rm n} {\bf{W}}^{\rm H}{\bf{W}}.
\end{multline}
From the above, the HB optimization problem {can be formulated as follows:}
\begin{equation}\label{qos_problem}
\begin{array}{l}
\mathop{\min}\limits_{\bf W_{\rm RF},W_{\rm BB},F_{\rm RF},F_{\rm BB}} \quad  {\rm Tr}\left(  {\bf{F}}_{\rm{BB}}^{\rm H}{\bf{F}}_{\rm{RF}}^{\rm H}{\bf{F}}_{\rm{RF}}{\bf{F}}_{\rm{BB}} \right) \\
\quad \text{subject to} \left\{
\begin{array}{l}
\hspace{-8 pt }\mathop{\max}\limits_{\| \boldsymbol\Delta\|_{\rm F}\le \varepsilon}{\bf E}_{k}({\bf F}_{\rm RF}{\bf F}_{\rm BB},{\bf W}_{\rm RF}{\bf W}_{\rm BB};{\boldsymbol \Delta})\le \rho_k, \forall k \\
{\bf F_{\rm RF}}\in \mathcal{F}_{\rm FC}/ \mathcal{F}_{\rm PC}, \quad {\bf W_{\rm RF}}\in \mathcal{W}_{\rm FC} /\mathcal{W}_{\rm PC},
\end{array}
\right.
\end{array}
\end{equation}
where $\rho_k$ is the maximum acceptable MSE of data stream $k$.

The problem \eqref{qos_problem} is non-convex due to {the} coupling between variables and the {erroneous CSI}. Hence, directly solving \eqref{qos_problem} is intractable. {To overcome this, we decouple the baseband beamforming from the RF beamforming,} and utilize a two-stage structure for both the baseband precoder and the baseband combiner. As a result, and as shown in Fig. \ref{hybrid_fig}, we have ${\bf F}_{\rm BB} = {\bf F}_{\rm BB}^{(2)} {\bf F}_{\rm BB}^{(1)}$ and ${\bf W}_{\rm BB} = {\bf W}_{\rm BB}^{(2)} {\bf W}_{\rm BB}^{(1)}$, and {we} obtain ${\bf F}_{\rm RF} {\bf F}_{\rm BB}^{(2)}$ and ${\bf W}_{\rm RF} {\bf W}_{\rm BB}^{(2)}$ as the RF precoder and combiner, respectively.

The eigenvectors of $\widehat {\bf H}$ are the optimal transmit and receive directions for robust sum MSE minimization \cite{wang2010robust, wang2011joint}. {Drawing on} this, we obtain {the} RF precoder and combiner from the right and left eigenvectors of $\widehat {\bf H}$, {respectively. We then find} ${\bf F}_{\rm BB}^{(1)}$ and ${\bf W}_{\rm BB}^{(1)}$ by solving \eqref{qos_problem} in such a way that the issues of infinitely many constraints and infeasibility are properly resolved.
\vspace{-12 pt}
\section{RF Beamforming}\label{analog_section}
{In} \cite{wang2010robust} {and} \cite{wang2011joint}, $\widehat {\bf H}$ is used in fully digital eigen-beamforming for min-max optimization {using the} worst-case approach, where {the} eigenvectors of $\widehat {\bf H}$ set the direction of {the} transmit and receive beams. We take a similar approach in our QoS-aware HB, but instead of minimizing {the} sum MSEs in \cite{wang2010robust, wang2011joint}, we satisfy {each} individual MSE constraint while adjusting {the} transmit power in each data stream.

We obtain ${\bf F}_{\rm RF}$ and ${\bf W}_{\rm RF}$ from the right and left eigenvectors of $\widehat {\bf H}$. The magnitude of each entry in ${\bf F}_{\rm RF}$ and ${\bf W}_{\rm RF}$ is equal to unity, which may not correspond to {the} eigenvectors of $\widehat {\bf H}$. As shown in \cite{tajalli2021qos}, only the phase in {the} eigenvectors of $\widehat {\bf H}$ is {used} to find ${\bf F}_{\rm RF}$ and ${\bf W}_{\rm RF}$, and the corresponding magnitudes are ignored, resulting in inaccurate eigenvectors. {To improve accuracy, we use ${\bf F}_{\rm BB}^{(2)}$ and ${\bf W}_{\rm BB}^{(2)}$ to obtain linear combinations of exponential entries in ${\bf F}_{\rm RF}$ and ${\bf W}_{\rm RF}$, and we approximate the right and left eigenvectors of $\widehat {\bf H}$ by ${\bf F}_{\rm RF} {\bf F}_{\rm BB}^{(2)}$ and ${\bf W}_{\rm RF} {\bf W}_{\rm BB}^{(2)}$, respectively.}

Let $\widehat {\bf H} = {\bf U}_{\rm H} {\boldsymbol \Sigma}_{\rm H} {\bf V}^{\rm H}_{\rm H}$ be the singular value decomposition (SVD) of $\widehat {\bf H}$, where ${\bf U}_{\rm H}$ and ${\bf V}_{\rm H}$ are the left and right eigenvectors of $\widehat {\bf H}$, respectively, and ${\boldsymbol \Sigma}_{\rm H}$ is a diagonal matrix with singular values of $\widehat {\bf H}$ in decreasing order. We obtain ${\bf F}_{\rm RF}{\bf F}_{\rm BB}^{(2)}$ and ${\bf W}_{\rm RF}{\bf W}_{\rm BB}^{(2)}$ as approximations of $\widetilde{\bf V}_{\rm H} = [{\bf V}_{\rm H}]_{1:N_{\rm s}}$ and $\widetilde{\bf U}_{\rm H} = [{\bf U}_{\rm H}]_{1:N_{\rm s}}$ in {a} Euclidean sense. Thus, {the} RF precoder is obtained by solving\vspace{-2 mm}
\begin{equation}\label{distance_precoder_problem}
\begin{array}{ccc}
& \mathop{\min}\limits_{\bf F_{\rm RF},F_{\rm BB}^{\rm (2)}} &   \|{\bf F}_{\rm RF} {\bf F}_{\rm BB}^{\rm (2)} - \widetilde{\bf V}_{\rm H} \|^2_{\rm F} \\
& \text{subject to} & {\bf F}_{\rm RF} \in \mathcal{F}_{\rm FC}/\mathcal{F}_{\rm PC}.
\end{array}
\end{equation}
{To obtain the RF combiner, we solve a similar problem}. Problem \eqref{distance_precoder_problem} is non-convex, and the alternating minimization method is used in \cite{tajalli2021qos} to solve \eqref{distance_precoder_problem} by finding the phase shifts in $\bf F_{\rm RF}$ for a given ${\bf F}_{\rm BB}^{\rm (2)}$ and vice versa until terminated. The phase shift $(m,l)$ in $\bf F_{\rm RF}$ is $\varphi_{m,l}^\star=-\angle \delta_{m,l}$, where 
\begin{equation}\label{rf_phase}
\delta_{k,l}=\left \{
\begin{array}{ll}
\sum\limits_{n=1}^{N_{\rm t}^{\rm RF}} \gamma_{m,n}^* [{\bf F}_{\rm BB}^{\rm (2)}]_{l,n}, & \text{in FCS}\\
\sum\limits_{n=1}^{N_{\rm t}^{\rm RF}} [\widetilde{{\bf V}}_{\rm W}]_{m,n}^* [{\bf F}_{\rm BB}^{\rm (2)}]_{l,n}, & \text{in PCS,}
\end{array}
\right.
\end{equation}
and $\gamma_{m,n} = [\widetilde{\bf V}_{\rm H}]_{m,n} - \sum \limits_{i \neq l} e^{j\varphi_{m,i}} [{\bf F}_{\rm BB}^{\rm (2)}]_{i,n}$ \cite{tajalli2021qos}.

Having obtained $\bf F_{\rm RF}$, the least square solution for ${\bf F}_{\rm BB}^{\rm (2)}$ is
\begin{equation}\label{least_precoder}
{\bf F}_{\rm BB}^{\rm (2)} = \left ({\bf F}_{\rm RF}^{\rm H} {\bf F}_{\rm RF}\right )^{-1} {\bf F}_{\rm RF}^{\rm H} \widetilde{\bf V}_{\rm H}.
\end{equation}
Similarly, ${\bf W}_{\rm RF}$ and ${\bf W}_{\rm BB}^{\rm (2)}$ can be obtained by replacing $\widetilde{\bf V}_{\rm H}$ with $\widetilde{\bf U}_{\rm H}$.

\section{Uncertainty Region of Effective Channel}\label{set_section}
Now that we have obtained ${\bf F}_{\rm RF}$ and ${\bf W}_{\rm RF}$, we need to find ${\bf F}_{\rm BB}^{\rm (1)}$ and ${\bf W}_{\rm BB}^{\rm (1)}$ for the worst-case channel in each data stream{; that is, we need to maximize the MSE} of data stream $k$ over $\|\boldsymbol\Delta\|_{\rm F}\le \varepsilon$. {The channel error matrix $\boldsymbol\Delta$ is $N_{\rm r} \times N_{\rm t}$ in size, and the MSE maximization over this large matrix is computationally intensive.} To mitigate this, instead of {the} channel's uncertainty region, we focus on the effective channel ${\bf H}_{\rm eff}={\bf W}_{\rm BB}^{\rm (2) H}{\bf W}_{\rm RF}^{\rm H} {\bf H}{\bf F_{\rm RF}}{{\bf F}_{\rm BB}^{(2)}}$ uncertainty region, {which is $N_{\rm s}\times N_{\rm s}$ in size.}

By plugging \eqref{channel_error} into the expression of the effective channel we have
\begin{equation}\label{effective_error}
{\bf H}_{\rm eff}={\bf W}_{\rm BB}^{\rm (2) H}{\bf W}_{\rm RF}^{\rm H} ({\widehat{\bf H}}+\boldsymbol\Delta){\bf F_{\rm RF}}{{\bf F}_{\rm BB}^{(2)}}={\widehat{\bf H}}_{\rm eff}+\boldsymbol\Delta_{\rm eff},
\end{equation}
where ${\widehat{\bf H}}_{\rm eff}={\bf W}_{\rm BB}^{\rm (2) H}{\bf W}_{\rm RF}^{\rm H} {\widehat{\bf H}}{\bf F_{\rm RF}}{{\bf F}_{\rm BB}^{(2)}}$ and $\boldsymbol\Delta_{\rm eff}={\bf W}_{\rm BB}^{\rm (2) H}{\bf W}_{\rm RF}^{\rm H} \boldsymbol\Delta{\bf F_{\rm RF}}{{\bf F}_{\rm BB}^{(2)}}$ {represent} the nominal effective channel and the effective channel error, respectively.

{We find the radius ${\varepsilon}_{\rm eff}$ of the effective channel's uncertainty region so that $\| \boldsymbol\Delta\|_{\rm F}\le \varepsilon$ gives $\|\boldsymbol\Delta_{\rm eff}\|_{\rm F} \le  {\varepsilon}_{\rm eff}$. Recall that $\varepsilon$ is set so that ${\rm Pr}\{\|\boldsymbol\Delta \|_{\rm F}\le \varepsilon\} =  P_{\rm in}$. Similarly, we find $\varepsilon_{\rm eff}$ so that ${\rm Pr}\{\|\boldsymbol\Delta_{\rm eff} \|_{\rm F}\le \varepsilon_{\rm eff}\} =  P_{\rm in}$; that is, we find $\varepsilon_{\rm eff}$ to guarantee QoS with probability $P_{\rm in}$. We obtain the probability distribution of $\|\boldsymbol\Delta_{\rm eff}\|^2_{\rm F}$ by} 
\begin{equation}
\|\boldsymbol\Delta_{\rm eff}\|_{\rm F}^2 ={\rm Tr}(\boldsymbol\Delta_{\rm eff}^{\rm H}\boldsymbol\Delta_{\rm eff})= {\rm vec}(\boldsymbol\Delta_{\rm eff})^{\rm H}{\rm vec}(\boldsymbol\Delta_{\rm eff}),
\end{equation}
where ${\rm vec}(\boldsymbol\Delta_{\rm eff})= \left({\bf F}_{\rm BB}^{(2)\rm T}{\bf F}_{\rm RF}^{\rm T}\otimes{\bf W}_{\rm BB}^{\rm (2) H}{\bf W}_{\rm RF}^{\rm H}\right) \boldsymbol\delta$ and $\boldsymbol\delta={\rm vec}(\boldsymbol\Delta)$. Therefore,
\begin{multline}\label{effective_norm}
\|\boldsymbol\Delta_{\rm eff}\|_{\rm F}^2 = \boldsymbol\delta^{\rm H} \Big(\underbrace{{\bf F}_{\rm RF}^*{\bf F}_{\rm BB}^{(2)*}{\bf F}_{\rm BB}^{(2)\rm T}{\bf F}_{\rm RF}^{\rm T}}_{{\bf A}_{\rm F}}\otimes\\
\underbrace{{\bf W}_{\rm RF}{\bf W}_{\rm BB}^{(2)}{\bf W}_{\rm BB}^{\rm (2) H}{\bf W}_{\rm RF}^{\rm H}}_{{\bf A}_{\rm W}}\Big) \boldsymbol\delta = \boldsymbol\delta^{\rm H} {\bf A}_{\rm RF} \boldsymbol\delta.
\end{multline}
The eigenvalue decomposition of $\bf A_{\rm RF}$ is ${\bf U}_{\rm RF} {\boldsymbol \Lambda}_{\rm RF} {\bf U}_{\rm RF}^{\rm H}$, where ${\bf U}_{\rm RF}$ and ${\boldsymbol \Lambda}_{\rm RF}$ include {the} eigenvectors and eigenvalues of ${\bf A}_{\rm RF}$, respectively. From \eqref{effective_norm}, we have
\begin{equation}\label{delta_eq}
\|\boldsymbol\Delta_{\rm eff}\|_{\rm F}^2 = \boldsymbol\delta^{\rm H} {\bf A}_{\rm RF} \boldsymbol\delta = \boldsymbol\delta^{\rm H} {\bf U}_{\rm RF} {\boldsymbol \Lambda}_{\rm RF} {\bf U}_{\rm RF}^{\rm H} \boldsymbol\delta = \tilde{\boldsymbol\delta}^{\rm H} {\boldsymbol \Lambda}_{\rm RF}  \tilde{\boldsymbol\delta},
\end{equation}
where $\tilde{\boldsymbol\delta} = {\bf U}_{\rm RF}^{\rm H} \boldsymbol\delta$. Rotating a vector by a unitary matrix does not change its statistics{; in other words, the} probability distributions of $\tilde{\boldsymbol\delta}$ and $\boldsymbol\delta$ are identical. In practice, {the} entries of $\boldsymbol\delta$ and $\tilde{\boldsymbol\delta}$ have {the} normal distribution \cite{cai2019robust}. Since ${\rm rank}({\bf A}_{\rm RF}) = N_{\rm s}^2$, ${\boldsymbol \Lambda}_{\rm RF}$ includes $N_{\rm s}^2$ non-zero eigenvalues. Hence,\vspace{-4 mm}
\begin{equation}\label{delta_sum}
\|\boldsymbol\Delta_{\rm eff}\|_{\rm F}^2 = \sum_{n=1}^{N_{\rm s}^2}\lambda_{{\rm RF},n}|\tilde{\delta}_n|^2,
\end{equation}
where $\lambda_{{\rm RF},n}$ is the $n^{\rm th}$ eigenvalue of ${\bf A}_{\rm RF}$, and $\tilde{\delta}_n$ is the $n^{\rm th}$ entry of $\tilde{\boldsymbol\delta}$. When {the} entries of $\boldsymbol\delta$ are i.i.d., $\|\boldsymbol\Delta_{\rm eff}\|_{\rm F}^2$ is a linear combination of ${N_{\rm s}^2}$ i.i.d. exponential random variables with individual means. Therefore, $\|\boldsymbol\Delta_{\rm eff}\|_{\rm F}^2$ is a hypoexponentially distributed random variable \cite{yanev2020exponential}. 

{To obtain the probability distribution of $\|\boldsymbol\Delta_{\rm eff}\|_{\rm F}^2$, we find the eigenvalues of ${\bf A}_{\rm RF}$. From }\eqref{effective_norm},{ the eigenvalues of ${\bf A}_{\rm RF}$ are obtained by multiplying the eigenvalues of ${{\bf A}_{\rm F}}$ and ${{\bf A}_{\rm W}}$. To avoid computing the SVD of ${{\bf A}_{\rm F}}$ and ${{\bf A}_{\rm W}}$, we note that ${\bf F}_{\rm RF}{\bf F}_{\rm BB}^{(2)}$ is an approximation of the semi-unitary matrix $\widetilde{\bf V}_{\rm H}$ in }\eqref{distance_precoder_problem}. {As shown in} \cite{tajalli2021qos}{, we assume that ${\bf F}_{\rm RF}{\bf F}_{\rm BB}^{(2)}$ is approximately semi-unitary; that is, ${\bf F}_{\rm RF}^{\rm H}{\bf F}_{\rm BB}^{\rm (2) H}{\bf F}_{\rm RF}{\bf F}_{\rm BB}^{(2)}\approx {\bf I}_{N_{\rm s}}$. The same is true for ${\bf W}_{\rm RF}{\bf W}_{\rm BB}^{(2)}$. Hence, we have}
\begin{multline}
\boldsymbol\lambda({\bf A}_{\rm RF}) =\\
 \boldsymbol\lambda({\bf F}_{\rm RF}^*{\bf F}_{\rm BB}^{(2)*}{\bf F}_{\rm BB}^{(2)\rm T}{\bf F}_{\rm RF}^{\rm T}) \otimes \boldsymbol\lambda({\bf W}_{\rm RF}{\bf W}_{\rm BB}^{(2)}{\bf W}_{\rm BB}^{\rm (2) H}{\bf W}_{\rm RF}^{\rm H})=\\
\boldsymbol\lambda({\bf F}_{\rm RF}^{\rm H}{\bf F}_{\rm BB}^{\rm (2) H}{\bf F}_{\rm RF}{\bf F}_{\rm BB}^{(2)}) \otimes \boldsymbol\lambda({\bf W}_{\rm RF}^{\rm H}{\bf W}_{\rm BB}^{\rm (2) H}{\bf W}_{\rm RF}{\bf W}_{\rm BB}^{(2)})\approx \\
{\bf 1}_{N_{\rm s}} \otimes {\bf 1}_{N_{\rm s}} = {\bf 1}_{N_{\rm s}^2},
\end{multline}
resulting in $\lambda_{{\rm RF},n} \approx 1, n=1,\ldots, N_{\rm s}^2$, which gives $\|\boldsymbol\Delta_{\rm eff}\|_{\rm F}^2 \approx \sum_{n=1}^{N_{\rm s}^2}|\tilde{\delta}_n|^2$. Note that $\boldsymbol \delta \sim \mathcal{CN}({\bf 0}, \sigma^2_{\rm e}{\bf I}_{N_{\rm t}N_{\rm r}})$ gives $\tilde{\boldsymbol\delta} \sim \mathcal{CN}({\bf 0}, \sigma^2_{\rm e}{\bf I}_{N_{\rm t}N_{\rm r}})$, because $ \tilde{\boldsymbol\delta}$ is obtained from a unitary rotation of  $\boldsymbol\delta$. As a result, $\sum_{n=1}^{N_{\rm s}^2}|\tilde{\delta}_n|^2$ is an Erlang distributed random variable with shape $ N_{\rm s}^2 $ and rate $\sigma^{-2}_{\rm e}$. Therefore, we have $\varepsilon_{\rm eff}^2=\phi_2^{-1}(P_{\rm in})$, where $\phi_2(\cdot)$ is the CDF of Erlang distribution with shape $ N_{\rm s}^2 $ and rate $\sigma^{-2}_{\rm e}$. {As mentioned in Section} \ref{CSI model}, $\varepsilon^2=\phi_1^{-1}(P_{\rm in})$. For two Erlang distributions with the same rate and different shapes $a_1$ and $a_2$, inequality $a_2 \ll a_1$ gives $\phi_2^{-1}(b) \ll \phi_1^{-1}(b)$ for $0<b<1$. Since $N_{\rm s}^2 \ll N_{\rm t}N_{\rm r}$, we have $\phi_2^{-1}(P_{\rm in}) \ll \phi_1^{-1}(P_{\rm in})$, which gives $\varepsilon_{\rm eff}^2 \ll \varepsilon^2$, i.e., the radius of {the} effective channel's uncertainty region is significantly smaller than the radius of {the} channel's uncertainty region, which we will numerically validate.
\vspace{-5 pt}
\section{Iterative Baseband Beamforming}\label{iterative_section}
In this section, we find ${\bf F}_{\rm BB}^{\rm (1) }$ and ${\bf W}_{\rm BB}^{\rm (1)}$. Having obtained ${\bf F}_{\rm RF}{\bf F}_{\rm BB}^{(2)}$, ${\bf W}_{\rm RF}{\bf W}_{\rm BB}^{(2)}$, and $\varepsilon_{\rm eff}$, the optimization problem \eqref{qos_problem} {can be reduced to}
\begin{equation}\label{qos_bb_problem}
\begin{array}{l}
\mathop{\min}\limits_{{\bf W}_{\rm BB}^{\rm (1)},{\bf F}_{\rm BB}^{\rm (1)}} \qquad \qquad {\rm Tr}\left(  {\bf{F}}_{\rm{BB}}^{\rm (1)H}{\bf A}_{\rm P}{\bf{F}}_{\rm BB}^{(1)} \right) \\
 \text{subject to}
\begin{array}{l}
\mathop{\max}\limits_{\| \boldsymbol\Delta_{\rm eff}\|_{\rm F}\le \varepsilon_{\rm eff}}{\bf E}_{k}({\bf F}_{\rm BB}^{\rm (1)},{\bf w}_{{\rm BB}, k}^{\rm (1)};{\boldsymbol \Delta}_{\rm eff})\le \rho_k, \forall k,
\end{array}
\end{array}
\end{equation}
where ${\bf A}_{\rm P} = {\bf F}_{\rm BB}^{\rm (2)H}{\bf{F}}_{\rm{RF}}^{\rm H}{\bf{F}}_{\rm{RF}}{\bf F}_{\rm BB}^{\rm (2)}$ and ${\bf w}_{{\rm BB}, k}^{\rm (1)}$ is the $k^{\rm th}$ column of ${\bf W}_{\rm BB}^{\rm (1)}$. From \eqref{mse_wc} and by some matrix manipulations, we get 
\begin{multline}\label{mse_bb}
{\bf E}_{k}({\bf F}_{\rm BB}^{\rm (1)},{\bf w}_{{\rm BB}, k}^{\rm (1)};{\boldsymbol \Delta}_{\rm eff}) = [\widehat{\bf E}]_{k,k} +\\
2\underbrace{{\rm Re}\left\{{\bf w}_{{\rm BB}, k}^{\rm (1)H}{\boldsymbol \Delta}_{\rm eff}{\bf F}_{\rm BB}^{\rm (1)}\left({\bf F}_{\rm BB}^{\rm (1)H}\widehat{\bf H}_{\rm eff}^{\rm H}{\bf w}_{{\rm BB}, k}^{\rm (1)}-{\bf i}_{k}\right)\right\}}_{[{\bf X}]_{k,k}}+\\
\underbrace{{\bf w}_{{\rm BB}, k}^{\rm (1)H}{\boldsymbol \Delta}_{\rm eff}{\bf F}_{\rm BB}^{\rm (1)}{\bf F}_{\rm BB}^{\rm (1)H}{\boldsymbol \Delta}_{\rm eff}^{\rm H}{\bf w}_{{\rm BB}, k}^{\rm (1)}}_{[{\bf Y}]_{k,k}},
\end{multline}
where ${\bf i}_{k}$ is the $k^{\rm th}$ column of ${\bf I}_{N_{\rm s}}$, and $[\widehat{\bf E}]_{k,k}$ is the MSE of data stream $k$ obtained from \eqref{mse_wc} when $\boldsymbol \Delta = \bf 0$.
\vspace{-2 mm}
\subsection{Feasibility Analysis}

When perfect CSI is available, it is possible to make $[{\hat{\bf E}}]_{k,k}$ arbitrarily small by consuming enough transmit power \cite{palomar2007mimo}. Hence, \eqref{qos_bb_problem} is feasible when $\boldsymbol \varepsilon_{\rm eff} = \bf 0$. {By} contrast, when CSI is imperfect or the highest acceptable MSE $\rho_k$ is small, there is a non-zero minimum achievable MSE, which may make \eqref{qos_bb_problem} infeasible. In such cases, we need to analyze {the} feasibility of \eqref{qos_bb_problem} and discard infeasible data streams.

To find infeasible data streams, we add a non-negative value $\alpha_k$ to each $\rho_k$ in \eqref{qos_bb_problem} {and} make a corresponding change in objective function, {which yields}\vspace{-5 pt}
\begin{equation}\label{feasible_problem}
\begin{array}{l}
\mathop{\min}\limits_{{\bf W}_{\rm BB}^{\rm (1)},{\bf F}_{\rm BB}^{\rm (1)},{\boldsymbol \alpha}} \qquad \gamma{\rm Tr}\left(  {\bf{F}}_{\rm{BB}}^{\rm (1)H}{\bf A}_{\rm P}{\bf{F}}_{\rm BB}^{(1)} \right) + (1-\gamma)\sum\limits_{k=1}^{N_{\rm s}}\alpha_k\\
\text{subject to}\left\{
\begin{array}{l}
\hspace{-8 pt} \mathop{\max}\limits_{\| \boldsymbol\Delta_{\rm eff}\|_{\rm F}\le \varepsilon_{\rm eff}}{\bf E}_{k}({\bf F}_{\rm BB}^{\rm (1)},{\bf w}_{{\rm BB}, k}^{\rm (1)};{\boldsymbol \Delta}_{\rm eff})\le \rho_k + \alpha_k, \forall k \\
\alpha_k\ge 0,  \forall k,
\end{array}
\right.
\end{array}
\end{equation}
where $0<\gamma<1$ is the weight of each term in the objective function. When \eqref{qos_bb_problem} is feasible, we have $\alpha_k=0, \forall k$. {By} contrast, when \eqref{qos_bb_problem} is infeasible, $\alpha_k>0$ because the MSE of data stream $k$ exceeds $\rho_k$ in an infeasible problem. As such, each data stream in \eqref{qos_bb_problem} that corresponds to a positive $\alpha_k$ is infeasible and discarded.

We employ the cutting-set method \cite{mutapcic2009cutting} to address infinitely many constraints in \eqref{feasible_problem}. The cutting-set method solves \eqref{feasible_problem} with a limited number of channel errors in the uncertainty region instead of considering all such {errors}. This latter problem in iteration $q$ of the cutting-set method is called the sample problem, {which is} defined as\vspace{-5 pt}
\begin{equation}\label{sampled_problem}
\begin{array}{ll}
\mathop{\min}\limits_{{\bf W}_{\rm BB}^{\rm (1)},{\bf F}_{\rm BB}^{\rm (1)},{\boldsymbol \alpha}} \qquad \quad  \gamma{\rm Tr}\left(  {\bf{F}}_{\rm{BB}}^{\rm (1)H}{\bf A}_{\rm P}{\bf{F}}_{\rm BB}^{(1)} \right) + (1-\gamma)\sum\limits_{k=1}^{N_{\rm s}}\alpha_k\\
 \text{subject to}\left\{
\begin{array}{l}
{\bf E}_{k}({\bf F}_{\rm BB}^{\rm (1)},{\bf W}_{\rm BB}^{\rm (1)};{\boldsymbol \Delta}_{{\rm eff},k}^{\rm wc})\le \rho_k + \alpha_k,\\
\qquad \qquad \qquad \qquad \qquad \forall k \text{ and } \forall{\boldsymbol \Delta}_{{\rm eff},k}^{\rm wc}\in \mathcal{D}_k^q\\
\alpha_k\ge 0, \quad \forall k,
\end{array}
\right.
\end{array}
\end{equation}
where ${\boldsymbol \Delta}_{{\rm eff},k}^{\rm wc}$ is {the} worst-case error of data stream $k$, and $\mathcal{D}_k^q$ is a set of {the} worst-case effective errors in iteration $q$. We solve \eqref{feasible_problem} by assuming perfect CSI{; that is, $\mathcal{D}_k^1= \{\bf 0\}$. We then obtain the} worst-case channel error matrix for each data stream and compare {the} worst-case MSE with {the} highest acceptable MSE for that data stream. When a violation occurs, the corresponding worst-case effective error is appended to $\mathcal{D}_k^q$. In this way, the solution in iteration $q+1$ does not violate {the} MSE constraint for {the} worst-case channel in iteration $q$. This procedure continues until no violation occurs.

The sample problem \eqref{sampled_problem} is non-convex due to {the} coupling between ${\bf F}_{\rm BB}^{\rm (1)}$ and ${\bf W}_{\rm BB}^{\rm (1)}$ in {the} MSE constraints. We adopt the alternating minimization method, which solves two convex sub-problems, one for ${\bf F}_{\rm BB}^{\rm (1)}$ and $\boldsymbol \alpha$, and another for ${\bf W}_{\rm BB}^{\rm (1)}$. The baseband precoding sub-problem is\vspace{-5 pt}
\begin{equation}\label{bbp_problem}
\begin{array}{l}
\mathop{\min}\limits_{{\bf F}_{\rm BB}^{\rm (1)},{\boldsymbol \alpha}} \qquad \quad  \gamma{\rm Tr}\left(  {\bf{F}}_{\rm{BB}}^{\rm (1)H}{\bf A}_{\rm P}{\bf{F}}_{\rm BB}^{(1)} \right) + (1-\gamma)\sum\limits_{k=1}^{N_{\rm s}}\alpha_k\\
 \text{subject to}\left\{
\begin{array}{l}
\hspace{-5 pt} {\bf E}_{k}({\bf F}_{\rm BB}^{\rm (1)};{\boldsymbol \Delta}_{{\rm eff},k}^{\rm wc})\le \rho_k + \alpha_k, \forall k \text{ and } \forall{\boldsymbol \Delta}_{{\rm eff},k}^{\rm wc}\in \mathcal{D}_k^q\\
\alpha_k\ge 0, \quad \forall k.
\end{array}
\right.
\end{array}
\end{equation}
After some matrix manipulations, \eqref{bbp_problem} is put in the standard QCQP form as in \eqref{qcqp_problem},
\begin{figure*}
\begin{equation}\label{qcqp_problem}
\begin{array}{cc}
\mathop{\min}\limits_{{\bf f}_{\rm BB}^{\rm (1)},{\boldsymbol \alpha}} &
\gamma{\bf f}_{\rm BB}^{\rm (1)H}({\bf I}_{N_{\rm s}}\otimes {\bf A}_{\rm P}){\bf f}_{\rm BB}^{\rm (1)}+ (1-\gamma)\sum\limits_{k=1}^{N_{\rm s}}\alpha_k =  g({\bf f }_{\rm BB}^{(1)},\boldsymbol\alpha)\\
\text{subject to}&\left\{
\begin{array}{l}
{\bf f}_{\rm BB}^{\rm (1)H}{\bf A}_{{\rm W},k} {\bf f}_{\rm BB}^{\rm (1)}-2{\rm Re}\left\{ {\bf b}_{{\rm W},k}^{\rm H}{\bf f}_{{\rm BB},k}^{\rm (1)} \right\}+c_k\le \rho_k + \alpha_k, \quad \forall k \text{ and } \forall{\boldsymbol \Delta}_{{\rm eff},k}^{\rm wc}\in \mathcal{D}_k^q\\
\alpha_k\ge 0, \quad \forall k,
\end{array}
\right.
\end{array}
\end{equation}
\hrulefill
\end{figure*}
where ${\bf f}_{\rm BB}^{\rm (1)} = {\rm vec}({\bf F }_{\rm BB}^{(1)})$,  ${\bf f}_{{\rm BB},k}^{\rm (1)}$ is the $k^{\rm th}$ column of ${\bf F }_{\rm BB}^{(1)}$, and
\begin{equation*}
\begin{array}{c}
{\bf A}_{{\rm W},k} = {\bf I}_{N_{\rm s}}\otimes {\bf H}_{{\rm eff},k}^{\rm wc H}{\bf w}_{{\rm BB}, k}^{\rm (1)}{\bf w}_{{\rm BB}, k}^{\rm (1)H}{\bf H}_{{\rm eff},k}^{\rm wc}, \qquad \\
{\bf b}_{{\rm W},k} = {\bf H}_{{\rm eff},k}^{\rm wc H}{\bf w}_{{\rm BB}, k}^{\rm (1)}, \qquad
c_k = {\bf w}_{{\rm BB}, k}^{\rm (1)H} {\bf R}_{\rm n}{\bf w}_{{\rm BB}, k}^{\rm (1)} + 1, \\
{\bf H}_{{\rm eff},k}^{\rm wc} = \widehat{\bf H}_{\rm eff} + {\boldsymbol \Delta}_{{\rm eff},k}^{\rm wc}, \text{and} \\
{\bf R}_{\rm n} = \sigma_{\rm n}^2{\bf W}_{\rm BB}^{\rm (2) H} {\bf W}_{\rm RF}^{\rm H}{\bf W}_{\rm RF}{\bf W}_{\rm BB}^{\rm (2)}.
\end{array}
\end{equation*}
{Here,} ${\bf A}_{\rm P}$ and ${\bf A}_{{\rm W},k}$ are positive semi-definite (PSD) matrices. Therefore, \eqref{qcqp_problem} is a convex QCQP, which can be efficiently solved by interior-point methods (IPMs).

Convergence of this method is guaranteed when ${\bf W}_{\rm BB}^{\rm (1)}$ does not shrink the feasible region of \eqref{sampled_problem}. To achieve a larger feasible region, for each $k$, we solve the following minimax problem\vspace{-2 mm}
\begin{equation}\label{minimax}
\min\limits_{{\bf w}_{{\rm BB},k}^{\rm (1)}}\quad \max\limits_{\forall {\boldsymbol \Delta}_{{\rm eff},k}^{\rm wc} \in \mathcal{D}_k^q} {\bf E}_{k}({\bf F}_{\rm BB}^{\rm (1)},{\bf w}_{{\rm BB},k}^{\rm (1)};{\boldsymbol \Delta}_{{\rm eff},k}^{\rm wc}).
\end{equation}
The standard QCQP form for the baseband combiner sub-problem is
\begin{equation}\label{bbc_problem}
\begin{array}{cc}
\mathop{\min}\limits_{{\bf w}_{{\rm BB},k}^{\rm (1)},\tau_k} & \tau_k\\
 \text{subject to}& {\bf w}_{{\rm BB},k}^{\rm (1)H}{\bf A}_{{\rm F},k} {\bf w}_{{\rm BB},k}^{\rm (1)}-\\
& 2{\rm Re}\left\{ {\bf b}_{{\rm F},k}^{\rm H}{\bf w}_{{\rm BB},k}^{\rm (1)} \right\}+1\le \tau_k, \quad \forall{\boldsymbol \Delta}_{{\rm eff},k}^{\rm wc}\in \mathcal{D}_k^q
\end{array}
\end{equation}
where $\tau_k$ is an auxiliary variable, ${\bf A}_{{\rm F},k} = {\bf H}_{{\rm eff},k}^{\rm wc}{\bf F}_{\rm BB}^{\rm (1)}{\bf F}_{\rm BB}^{\rm (1)H}{\bf H}_{{\rm eff},k}^{\rm wc H}+{\bf R}_{\rm n},$ and ${\bf b}_{{\rm F},k} = {\bf H}_{{\rm eff},k}^{\rm wc}{\bf f}_{{\rm BB}, k}^{\rm (1)}$. Note that ${\bf A}_{{\rm F},k}$ is PSD. Therefore, \eqref{bbc_problem} is a convex QCQP and can be efficiently solved by IPMs. The alternating minimization method in Algorithm \ref{sampled_algorithm} solves \eqref{bbp_problem} for a given ${\bf W}_{\rm BB}^{\rm (1)}$ and then solves \eqref{bbc_problem} for a given ${\bf F}_{\rm BB}^{\rm (1)}$ until convergence.

\begin{algorithm}[t!]
\begin{small}
	\label{sampled_algorithm}
	\DontPrintSemicolon
	\KwInput
	{
		Random ${\bf{W}}_{\rm{BB}}^{(1)(0)}, \epsilon, N_{\rm it}, r=0$
	}
	\Repeat{ ${\bf g}({\bf f}_{\rm BB}^{(1)(r-1)},\boldsymbol\alpha^{(r-1)})-{\bf g}({\bf f}_{\rm BB}^{(1)(r)},\boldsymbol\alpha^{(r)})\le \epsilon$ \rm or $r \ge N_{\rm it}$}
	{			
		$r=r+1$\\
		Solve \eqref{qcqp_problem} and return ${\bf F }_{\rm BB}^{(1)(r)}$ and $\boldsymbol\alpha^{(r)}$\\
		\For{$k=1$ \rm{to} $N_{\rm s}$}
		{
		Solve \eqref{bbc_problem} and return ${\bf w}_{{\rm BB},k}^{(1)(r)}$
		}
	}
	\KwOutput{$ {\bf F}_{\rm BB}^{(1)}$, $ {\bf W}_{\rm BB}^{(1)}$, and $\boldsymbol\alpha$}
	
	\caption{Solving \eqref{sampled_problem}}
\end{small}
\end{algorithm}
For a feasible \eqref{sampled_problem}, we have $\sum_{k=1}^{N_{\rm s}}\alpha_k = 0$. Otherwise, we need to eliminate infeasible data streams. To do so, we find the data stream that has the highest value of $\alpha_k$, {which is} ${\tilde k}=\argmax\limits_k \boldsymbol \alpha$. {We then eliminate the data stream ${\tilde k}$ and its MSE constraint,} and solve \eqref{sampled_problem} for the remaining data streams. This procedure is repeated until $\sum_{k=1}^{N_{\rm s}}\alpha_k = 0$.
\vspace{-3 mm}

\subsection{Worst-Case Analysis}\label{wc_su_subsection}

We find the worst-case channel for each feasible data stream, i.e., the effective error matrix that maximizes {the} MSE of each individual data stream. From \eqref{mse_bb}, {the} MSE of data stream $k$ is
\begin{equation}\label{mse_delta}
{\bf E}_{k}({\bf F}_{\rm BB}^{\rm (1)},{\bf w}_{{\rm BB}, k}^{\rm (1)};{\boldsymbol \Delta}_{\rm eff}) =
{\boldsymbol \delta}_{\rm eff}^{\rm H} {\bf A}_{\delta,k} {\boldsymbol \delta}_{\rm eff} + 
2{\rm Re}\{ {\bf b}_{\delta,k}^{\rm H}{\boldsymbol \delta}_{\rm eff}\} + [\widehat{\bf E}]_{k,k},
\end{equation}
where ${\boldsymbol \delta}_{\rm eff} = {\rm vec}({\boldsymbol \Delta}_{\rm eff})$, \begin{equation*}
\begin{array}{c}
{\bf A}_{\delta,k} ={\bf F}_{\rm BB}^{\rm (1)*}{\bf F}_{\rm BB}^{\rm (1)T} \otimes {\bf w}_{{\rm BB}, k}^{(1)}{\bf w}_{{\rm BB}, k}^{(1)\rm H}, \\
{\bf b}_{\delta,k} = ({\bf F}_{\rm BB}^{\rm (1)}{\bf F}_{\rm BB}^{\rm (1)H} \widehat{\bf H}_{\rm eff}^{\rm H}{\bf w}_{{\rm BB}, k}^{(1)}-{\bf f}^{(1)}_{{\rm BB}, k})^* \otimes {\bf w}_{{\rm BB}, k}^{(1)}.
\end{array}
\end{equation*}
The worst-case channel is the solution to the following {problem:}
\begin{equation} \label{wc_problem}
\begin{array}{ccc}
{\boldsymbol \delta}_{{\rm eff},k}^{\rm wc} =& \mathop{\argmax\limits_{{\boldsymbol \delta}_{\rm eff}}}&{\boldsymbol \delta}_{\rm eff}^{\rm H} {\bf A}_{\delta,k} {\boldsymbol \delta}_{\rm eff} + 
2{\rm Re}\{ {\bf b}_{\delta,k}^{\rm H}{\boldsymbol \delta}_{\rm eff}\}\\
&\text{subject to}&\boldsymbol\delta_{\rm eff}^{\rm H}\boldsymbol\delta_{\rm eff}\le \varepsilon_{\rm eff}^2,
\end{array}
\end{equation}
which is a convex problem since ${\bf A}_{\delta,k}$ is PSD. The Lagrangian function of \eqref{wc_problem} is
\begin{equation}
{\mathcal L}(\boldsymbol\delta_{\rm eff},\nu) = {\boldsymbol \delta}_{\rm eff}^{\rm H} {\bf A}_{\delta,k} {\boldsymbol \delta}_{\rm eff} + 2{\rm Re}\{ {\bf b}_{\delta,k}^{\rm H}{\boldsymbol \delta}_{\rm eff}\}
+\nu (\boldsymbol\delta_{\rm eff}^{\rm H}\boldsymbol\delta_{\rm eff}- \varepsilon_{\rm eff}^2),
\end{equation}
where $\nu \ge 0$ is the Lagrange multiplier. The maximizer of {the} Lagrangian function is\vspace{-2 mm}
\begin{equation}\label{wc_error}
{\boldsymbol \delta}_{{\rm eff},k}^{\rm wc} = - \left({\bf A}_{\delta,k}+\nu {\bf I}_{N_{\rm s}}\right)^{-1} {\bf b}_{\delta,k}.
\end{equation}
Using {the} Schur complement \cite{boyd2004convex}, the Lagrange multiplier is the solution to\vspace{-2 mm}
\begin{equation} \label{dual_problem}
\begin{array}{ccc}
& \mathop{\min\limits_{\psi,\nu}}&\psi\\
& \text{subject to}&\left\{
\begin{array}{l}
\nu\ge 0,\\
\left[
\begin{array}{cc}
{\bf A}_{\delta,k}+\nu {\bf I}_{N_{\rm s}} &  {\bf b}_{\delta,k}\\
{\bf b}_{\delta,k}^{\rm H} & \psi- \varepsilon_{\rm eff}^2
\end{array}
\right]\ge 0,
\end{array}
\right.
\end{array}
\end{equation}
where $\psi$ is an auxiliary variable, and $\bf A\ge 0$ ensures the positive semi-definiteness of $\bf A$. {Now,} \eqref{dual_problem} can be efficiently solved via the SDP-based method by standard solvers.

When {the} worst-case error in \eqref{wc_error} violates an MSE constraint, {such as} ${\bf E}_{k}({\bf F}_{\rm BB}^{\rm (1)},{\bf w}_{{\rm BB}, k}^{\rm (1)};{\boldsymbol \Delta}^{\rm wc}_{{\rm eff},k}) > \rho_k$, a new MSE constraint corresponding to ${\boldsymbol \Delta}^{\rm wc}_{{\rm eff},k}$ is added to \eqref{sampled_problem}. As a result, in the next iteration, we have ${\bf E}_{k}({\bf F}_{\rm BB}^{\rm (1)},{\bf w}_{{\rm BB}, k}^{\rm (1)};{\boldsymbol \Delta}^{\rm wc}_{{\rm eff},k}) \le \rho_k$. This process continues until no violation occurs.
\vspace{-5 pt}
\subsection{Summary}

\begin{figure}[t]
\centering
\includegraphics[width=8 cm]{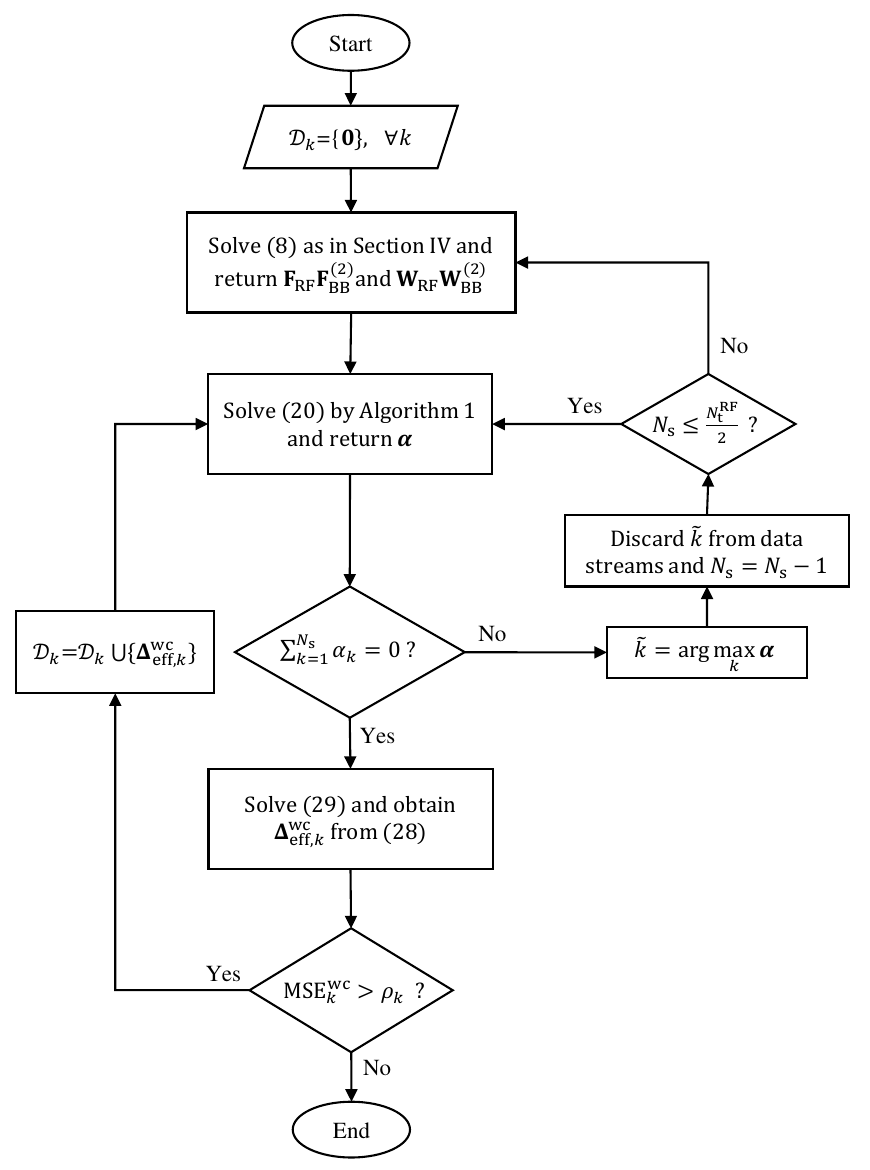}
\caption{{The flowchart of our iterative scheme for solving} \eqref{qos_problem}.}
\label{flowchart2}
\end{figure}

{To sum up our scheme, as shown in Fig.} \ref{flowchart2}{, we start with RF beamforming as discussed in Section} \ref{analog_section}{, and then we solve the sample problem} \eqref{sampled_problem} {via Algorithm} \ref{sampled_algorithm}. When \eqref{qos_bb_problem} is infeasible, at least one entry of $\boldsymbol \alpha$ is positive, and we discard the data stream with the highest MSE violation (i.e., the data stream that corresponds to the maximum entry in $\boldsymbol\alpha$). For a fixed number of RF chains, the performance of RF beamforming is improved (the Euclidean distance in \eqref{distance_precoder_problem} is reduced) by reducing the number of data streams. We re-do {the} RF beamforming when a data stream is discarded. RF beamforming is optimal, i.e., the objective of \eqref{distance_precoder_problem} is zero, when $N_{\rm s}\le \frac{N_{\rm t}^{\rm RF}}{2}$ \cite{tajalli2021qos}. It is not necessary to re-do {the} RF beamforming when $N_{\rm s}< \frac{N_{\rm t}^{\rm RF}}{2}$.

Algorithm \ref{sampled_algorithm} finds ${\bf F}_{\rm BB}^{\rm (1)}$, ${\bf W}_{\rm BB}^{\rm (1)}$, and $\boldsymbol\alpha$ by solving \eqref{sampled_problem}. When $\sum_{k=1}^{N_{\rm s}}\alpha_k=0$, we check if the solution satisfies MSE constraints for all {errors} in the effective channel's uncertainty region. To do so, when {the} MSE constraint for a data stream is violated, the error in the effective channel for that data stream is appended to the set of worst-case effective errors. Otherwise, the solution to \eqref{sampled_problem} satisfies MSE constraints for all norm-bounded effective errors and a sub-optimal solution to \eqref{qos_bb_problem} is obtained.

\vspace{-5 mm}
\subsection{Convergence}

{The convergence} of our RF beamforming in Section \ref{analog_section} is proved in \cite{tajalli2021qos}. In iteration $r$ of Algorithm \ref{sampled_algorithm}, the solution to \eqref{bbp_problem} minimizes the objective function of \eqref{sampled_problem}, and the solution to \eqref{bbc_problem} does not shrink the feasible region of \eqref{sampled_problem}. Hence, the objective function of \eqref{sampled_problem} in iteration $(r+1)$ is equal to or less than its value in iteration $r$, i.e., $g({\bf f}_{\rm BB}^{(1)(r+1)},\boldsymbol\alpha^{(r+1)})\le g({\bf f}_{\rm BB}^{(1)(r)},\boldsymbol\alpha^{(r)})$. Algorithm \ref{sampled_algorithm} is convergent because the objective function of \eqref{sampled_problem} is lower bounded to 0.

Note that discarding a data stream expands the feasible region of \eqref{sampled_problem}, {which means that the objective function of} \eqref{sampled_problem} becomes smaller, {although this does not impact} the convergence of our iterative scheme. {We should add} that appending a new error value to the set of worst-case errors in the effective channel does not impact the convergence of {the} cutting-set method \cite{mutapcic2009cutting}.
\vspace{-4 mm}
\subsection{Computational Complexity (CC)}

{The} CC of our iterative method is dominated by {the} CC of solving \eqref{sampled_problem} and \eqref{dual_problem}, and can be {done} via the well-known solver CVX \cite{grant2014cvx}, which utilizes IPMs. Algorithm \ref{sampled_algorithm} solves \eqref{sampled_problem} by iteratively solving \eqref{qcqp_problem} and \eqref{bbc_problem}, both of which are second order cone programming (SOCP) problems. From \cite[Section~6.6]{ben2001lectures}, {the} CC of utilizing IPM for solving a SOCP with $n$ variables and $m$ constraints is proportional to $n\sqrt{m+1}(n^2+m+mn^2)$. The optimization problem \eqref{qcqp_problem} includes $n=N_{\rm s}^2+N_{\rm s}$ variables and at most $m=N_{\rm max}N_{\rm s}$ constraints, where $N_{\rm max}$ is the maximum number of iterations in the cutting-set method. Therefore, {the} CC of solving \eqref{qcqp_problem} is ${\mathcal O}\left( N_{\rm max}^{1.5} N_{\rm s}^{7.5}\right)$. The optimization problem \eqref{bbc_problem} includes $n=N_{\rm s}+1$ variables and $m= N_{\rm max}$ constraints. {The problem is solved} $N_{\rm s}$ times for all columns of ${\bf W}_{\rm BB}^{(1)}$, which results in {the} CC of ${\mathcal O}\left\{ N_{\rm max}^{1.5} N_{\rm s}^{4}\right\}$. As a result, {the} CC of solving \eqref{sampled_problem} is dominated by {the} CC of \eqref{qcqp_problem}. Note that \eqref{qcqp_problem} is solved at most $N_{\rm it}$ times in Algorithm \ref{sampled_algorithm}, and {the} CC of solving \eqref{sampled_problem} is ${\mathcal O}\left( N_{\rm it} N_{\rm max}^{1.5} N_{\rm s}^{7.5}\right)$.

From \cite[Section~6.6]{ben2001lectures}, {the} CC of {using} IPM for solving SDP with $n$ variables and PSD matrices of size $m\times m$ is proportional to $n\sqrt{m+1}(n^2+nm^2+m^3)$. Hence, {the} CC of the SDP for solving \eqref{dual_problem} is ${\mathcal O}\left( N_{\rm s}^{6.5}\right)$, which is dominated by {the} CC of solving \eqref{sampled_problem}. The cutting-set method solves \eqref{sampled_problem} at most $N_{\rm max}$ times, and the overall CC of the iterative {scheme} is ${\mathcal O}\left( N_{\rm it} N_{\rm max}^{2.5} N_{\rm s}^{7.5}\right)$.
\vspace{-5 pt}
\section{Low-Complexity Baseband Beamforming}\label{low_section}

Our scheme in Section \ref{iterative_section} includes nested alternating minimization in the cutting-set method, which is computationally intensive. We now introduce a low-complexity method for designing the baseband precoder and combiner. {Drawing on} HB in the perfect CSI scenario \cite{tajalli2021qos}, we diagonalize the nominal effective channel in \eqref{effective_error} for baseband precoding {and use} MMSE combining in the baseband combiner. Next, we find the worst-case effective channel and determine the transmit power so that all MSE constraints for worst-case effective channels are satisfied.
\subsection{Baseband Precoder and Combiner}

As in \cite{tajalli2021qos}, we diagonalize the nominal effective channel {for} when ${\bf F}_{\rm BB}^{(1)} = {\bf V}{\bf P}^{\frac{1}{2}}$, where ${\bf V}$ includes the right singular vectors of ${\bf R}_{\rm n}^{-\frac{1}{2}}\widehat{\bf H}_{\rm eff}= {\bf U} {\boldsymbol \Sigma} {\bf V}^{\rm H}$, ${\bf U}$ includes the left singular vectors of ${\bf R}_{\rm n}^{-\frac{1}{2}}\widehat{\bf H}_{\rm eff}$, $\boldsymbol \Sigma = {\rm diag}\{\sigma_1,\ldots,\sigma_{N_{\rm s}}\}$ includes the singular values of ${\bf R}_{\rm n}^{-\frac{1}{2}}\widehat{\bf H}_{\rm eff}$ in decreasing order, and ${\bf P}={\rm diag}\{p_1,\ldots,p_{N_{\rm s}}\}$ is the power allocation matrix. Given ${\bf F}_{\rm BB}^{(1)}$, the MMSE baseband combiner is 
\begin{equation}\label{bb_combining}
{\bf W}_{\rm BB}^{(1)} = \left( \widehat{\bf H}_{\rm eff}{\bf F}_{\rm BB}^{(1)}{\bf F}_{\rm BB}^{\rm (1)H} \widehat{\bf H}_{\rm eff}^{\rm H} + {\bf R}_{\rm n}\right)^{-1} \widehat{\bf H}_{\rm eff}{\bf F}_{\rm BB}^{(1)}.\vspace{-5 mm}
\end{equation}
\subsection{Worst-Case Analysis}

With ${\bf F}_{\rm BB}^{(1)}$ and ${\bf W}_{\rm BB}^{(1)}$ as above, solving \eqref{qos_bb_problem} {can be reduced} to finding ${\bf P}$. {For this}, we first find {the} worst-case MSE in \eqref{mse_bb}. From the matrix inversion lemma \cite{horn2012matrix}, \eqref{bb_combining} is\vspace{-5 pt}
\begin{multline}\label{bb_inversion}
{\bf W}_{\rm BB}^{(1)} = \bigg[{\bf R}_{\rm n}^{-1} - {\bf R}_{\rm n}^{-1} \widehat{\bf H}_{\rm eff}{\bf F}_{\rm BB}^{(1)}
\Big( {\bf I}_{N_{\rm s}} + \\
\left. \left. {\bf F}_{\rm BB}^{\rm (1)H} \widehat{\bf H}_{\rm eff}^{\rm H} {\bf R}_{\rm n}^{-1} \widehat{\bf H}_{\rm eff}{\bf F}_{\rm BB}^{(1)} \right)^{-1} {\bf F}_{\rm BB}^{\rm (1)H} \widehat{\bf H}_{\rm eff}^{\rm H} {\bf R}_{\rm n}^{-1}\right] \widehat{\bf H}_{\rm eff}{\bf F}_{\rm BB}^{(1)}.\vspace{-4 mm}
\end{multline}
{Plugging ${\bf R}_{\rm n}^{-\frac{1}{2}}\widehat{\bf H}_{\rm eff}{\bf F}_{\rm BB}^{\rm (1)} = {\bf U} {\boldsymbol \Sigma} {\bf P}^{\frac{1}{2}}$ into} \eqref{bb_inversion}{ and noting}
\begin{align}\label{delta_simple}
{\bf W}_{\rm BB}^{\rm (1)H} {\boldsymbol \Delta}_{\rm eff}{\bf F}_{\rm BB}^{\rm (1)}= & 
\left( {\bf I}_{N_{\rm s}}+ {\boldsymbol \Sigma}^2{\bf P} \right)^{-1} {\bf P}^{\frac{1}{2}} {\boldsymbol \Sigma} {\bf U}^{\rm H}{\bf R}_{\rm n}^{-\frac{1}{2}}{\boldsymbol \Delta}_{\rm eff} {\bf V}{\bf P}^{\frac{1}{2}},\\
{\bf F}_{\rm BB}^{\rm (1)H} \widehat{\bf H}_{\rm eff}^{\rm H}{\bf W}_{\rm BB}^{\rm (1)} = &
{\bf P}{\boldsymbol \Sigma}^2 \left( {\bf I}_{N_{\rm s}} + {\boldsymbol \Sigma}^2{\bf P} \right)^{-1},
\end{align}
{gives $[{\bf X}]_{k,k}$ in }\eqref{mse_bb} as\vspace{-2 mm}
\begin{equation}\label{simple_x}
[{\bf X}]_{k,k} = \frac{\sigma_k  p_k}{(1+\sigma_k^2  p_k)^2} {\rm Re}\left\{ {\bf u}^{\rm H}_k {\bf R}_{\rm n}^{-\frac{1}{2}}{\boldsymbol \Delta}_{\rm eff} {\bf v}_k \right\},
\end{equation}
where ${\bf v}_k $ and ${\bf u}_k $ are columns $k$ of $\bf V$ and $\bf U$, respectively. {Using the} Cauchy-Schwarz inequality, the maximum value of $[{\bf X}]_{k,k}$ is obtained by setting
\begin{equation}\label{wc_x}
{\boldsymbol \Delta}_{{\rm eff},k}^{\rm wc,x} = \frac{\varepsilon_{\rm eff}} {\beta_k} {\bf R}_{\rm n}^{-\frac{1}{2}}{\bf u}_k {\bf v}^{\rm H}_k,
\end{equation}
where $\beta_k = \sqrt{{\bf u}^{\rm H}_k {\bf R}_{\rm n}^{-1} {\bf u}_k}$ is a coefficient defined such that $\| {\boldsymbol \Delta}_{{\rm eff},k}^{\rm wc,x}\|_{\rm F} = \varepsilon_{\rm eff}$. Substituting \eqref{wc_x} into \eqref{simple_x} gives the maximum value of $[{\bf X}]_{k,k}$ as
\begin{equation}\label{x_wc}
[{\bf X}^{\rm wc}]_{k,k} = \frac{\sigma_k  p_k}{(1+\sigma_k^2  p_k)^2} \beta_k \varepsilon_{\rm eff}.
\end{equation}

{From} \eqref{delta_simple}{, we get $[{\bf Y}]_{k,k}$ in }\eqref{mse_bb} as
\begin{align*}
[{\bf Y}]_{k,k} & =  \frac{\sigma_k^2  p_k}{(1+\sigma_k^2  p_k)^2} {\bf u}_k^{\rm H}{\bf R}_{\rm n}^{-\frac{1}{2}}{\boldsymbol \Delta}_{\rm eff} {\bf V}{\bf P}{\bf V}^{\rm H}{\boldsymbol \Delta}_{\rm eff}^{\rm H} {\bf R}_{\rm n}^{-\frac{1}{2}}{\bf u}_k \\
& = \frac{\sigma_k^2  p_k}{(1+\sigma_k^2  p_k)^2} \sum_{j=1}^{N_{\rm s}} \left| {\bf u}_k^{\rm H}{\bf R}_{\rm n}^{-\frac{1}{2}}{\boldsymbol \Delta}_{\rm eff} {\bf v}_j \right|^2 p_j \\
& \le \frac{\sigma_k^2  p_k}{(1+\sigma_k^2  p_k)^2} \sum_{j=1}^{N_{\rm s}} \left| {\bf u}_k^{\rm H}{\bf R}_{\rm n}^{-\frac{1}{2}}{\boldsymbol \Delta}_{\rm eff} {\bf v}_j \right|^2 p_{\rm max},\\
& = \frac{\sigma_k^2  p_k}{(1+\sigma_k^2  p_k)^2}p_{\rm max}{\boldsymbol \delta}_{\rm eff}^{\rm H} {\bf A}_{{\rm Y},{k}} {\boldsymbol \delta}_{\rm eff}
\end{align*}
where $p_{\rm max} = \max\limits_k p_k$, and ${\bf A}_{{\rm Y},{k}} = {\bf I}_{N_{\rm s}} \otimes {\bf R}_{\rm n}^{-\frac{1}{2}}{\bf u}_k{\bf u}_k^{\rm H}{\bf R}_{\rm n}^{-\frac{1}{2}}$. The maximum value of $[{\bf Y}]_{k,k}$ is obtained by solving\vspace{-2 mm}
\begin{equation} \label{y_wc_prob}
\begin{array}{ccc}
& \mathop{\max\limits_{{\boldsymbol \delta}_{\rm eff}}}&{\boldsymbol \delta}_{\rm eff}^{\rm H} {\bf A}_{{\rm Y},{k}} {\boldsymbol \delta}_{\rm eff} \\
&\text{subject to}&\boldsymbol\delta_{\rm eff}^{\rm H}\boldsymbol\delta_{\rm eff}\le \varepsilon_{\rm eff}^2,
\end{array}
\end{equation}
whose optimal solution is ${\boldsymbol \delta}_{{\rm eff},k}^{\rm wc,y} = \varepsilon_{\rm eff}{\bf u}_{\rm A,max}$, where ${\bf u}_{\rm A,max}$ is the eigenvector corresponding to the highest eigenvalue of ${\bf A}_{{\rm Y},{k}}$. The optimal value of \eqref{y_wc_prob} is $\varepsilon_{\rm eff}^2 \lambda_{\rm A,max}$, where $\lambda_{\rm A,max}$ is the highest eigenvalue of ${\bf A}_{{\rm Y},{k}}$, that is\vspace{-2 mm}
\begin{equation}
\lambda_{\rm A,max} = \lambda_{\rm max}\left( {\bf I}_{N_{\rm s}} \right) \lambda_{\rm max}\left( {\bf R}_{\rm n}^{-\frac{1}{2}}{\bf u}_k{\bf u}_k^{\rm H}{\bf R}_{\rm n}^{-\frac{1}{2}} \right),
\end{equation}
where we use $\lambda_{\rm max}(\bf A \otimes B) = \lambda_{\rm max}(\bf A)\lambda_{\rm max}( B)$. {Here,} ${\bf R}_{\rm n}^{-\frac{1}{2}}{\bf u}_k{\bf u}_k^{\rm H}{\bf R}_{\rm n}^{-\frac{1}{2}}$ is rank-one and its eigenvector is $\frac{1}{\beta_k}{\bf R}_{\rm n}^{-\frac{1}{2}}{\bf u}_k$. Hence, $\lambda_{\rm max}\left( {\bf R}_{\rm n}^{-\frac{1}{2}}{\bf u}_k{\bf u}_k^{\rm H}{\bf R}_{\rm n}^{-\frac{1}{2}} \right)  = \beta_k^2$, which gives $\lambda_{\rm A,max} =\beta_k^2$ and 
\begin{equation}\label{y_wc}
[{\bf Y}^{\rm wc}]_{k,k} = \frac{\sigma_k^2  p_k p_{\rm max}}{(1+\sigma_k^2  p_k)^2}\varepsilon_{\rm eff}^2\beta_k^2.
\end{equation}
From \eqref{x_wc} and \eqref{y_wc}, {the} worst-case MSE in \eqref{mse_bb} is
\begin{multline}\label{wc_mse}
[{\bf E}^{\rm wc}]_{k,k} = \frac{1}{1+\sigma_k^2 p_k} + 2\frac{\sigma_k  p_k}{(1+\sigma_k^2  p_k)^2} \beta_k \varepsilon_{\rm eff} + \\
\frac{\sigma_k^2  p_k p_{\rm max}}{(1+\sigma_k^2  p_k)^2}\varepsilon_{\rm eff}^2\beta_k^2,
\end{multline}
where $[\widehat{\bf E}]_{k,k} =  \frac{1}{1+\sigma_k^2 p_k}$ \cite{tajalli2021qos}.
\vspace{-2 mm}
\subsection{Feasibility Analysis}

Here we obtain a feasible subset of data streams. A data stream is feasible when {the transmit power satisfies the worst-case MSE for that stream.} Note that $[{\bf E}^{\rm wc}]_{k,k}$ is a decreasing function for large values of $p_k$. Therefore, the minimum of $[{\bf E}^{\rm wc}]_{k,k}$ is\vspace{-5 pt}
\begin{equation}
{\lim}_{p_k \to +\infty} [{\bf E}^{\rm wc}]_{k,k} = \frac{1}{\sigma^2_k}\varepsilon_{\rm eff}^2\beta_k^2,
\end{equation}
which gives a sufficient condition for feasibility of \eqref{qos_bb_problem} as $\frac{1}{\sigma^2_k}\varepsilon_{\rm eff}^2\beta_k^2\le \rho_k$. Hence, small values of $\sigma_k$ or large values of $\varepsilon_{\rm eff}$	may violate the MSE constraint in stream $k$, causing infeasibility of \eqref{qos_bb_problem}.

From Section \ref{set_section}, we have ${\bf W}_{\rm RF}^{\rm H}{\bf W}_{\rm BB}^{\rm (2) H}{\bf W}_{\rm RF}{\bf W}_{\rm BB}^{(2)} \approx {\bf I}_{N_{\rm s}}$. Therefore, $\beta_k^2 \approx \frac{1}{\sigma^2_{\rm n}}$ and $\sigma^2_k \approx \frac{\widehat{\sigma}^2_k}{\sigma^2_{\rm n}}$, where $\widehat{\sigma}_k$ is the $k^{\rm th}$ singular value of $\widehat{\bf H}_{\rm eff}$. As a result, the sufficient condition is simplified to $\frac{1}{\widehat{\sigma}^2_k}\varepsilon_{\rm eff}^2 \le \rho_k$. For a given number of RF chains, the singular values of $\widehat{\bf H}_{\rm eff}$ are larger for smaller $N_{\rm s}$ \cite{tajalli2021qos}, which means discarding an infeasible data stream may increase ${\widehat{\sigma}^2_k}$, resulting in a higher probability for other data streams to be feasible. Similar to the iterative method, we re-do {the} RF beamforming when an infeasible data stream with the highest value of $\frac{1}{\sigma^2_k}\varepsilon_{\rm eff}^2\beta_k^2- \rho_k$ is discarded.\vspace{-10 pt}
\subsection{Transmit Power Allocation}
Having obtained {the} worst-case MSE in \eqref{wc_mse}, problem \eqref{qos_bb_problem} {can be reduced to the following:}
\begin{equation}\label{power_problem}
\begin{array}{ccc}
& \mathop{\min}\limits_{\{p_1,\ldots p_{N_{\rm s}} \}} & \sum\limits_{k=1}^{N_{\rm s}} p_k\\
& \text{subject to}&
\begin{array}{l}
\frac{1}{1+\sigma_k^2 p_k} + 2\frac{\sigma_k  p_k}{(1+\sigma_k^2  p_k)^2} \beta_k \varepsilon_{\rm eff} +\\
 \frac{\sigma_k^2  p_k p_{\rm max}}{(1+\sigma_k^2  p_k)^2}\varepsilon_{\rm eff}^2\beta_k^2\le \rho_k, \quad \forall k,
\end{array}
\end{array}
\end{equation}
where ${\bf A}_{\rm P} \approx {\bf I}_{N_s}$. In \eqref{power_problem}, $p_{\rm max}$ is coupled to $p_k$. To remove the coupling, we first find $p_{\rm max}${, which satisfies} the constraint of data stream $k_{\rm max}$, followed by finding {the} transmit power levels of the remaining streams. However, the index $k_{\rm max}$ is not known prior to solving \eqref{power_problem}, but we presume {that} it corresponds to the same data stream in the perfect CSI scenario. Hence, the index $k_{\rm max}$ can be obtained by solving the nominal optimization problem prior to solving \eqref{power_problem}.

From \cite{tajalli2021qos}, in the {scenario with perfect CSI, the} transmit power in data stream $k$ is $ \frac{1}{\sigma_k^2}\left( \frac{1}{\rho_k}-1 \right)$. Hence, $k_{\rm max} = \argmax\limits_{k} \frac{1}{\sigma_k^2}\left( \frac{1}{\rho_k}-1 \right)$. The worst-case MSE constraint in \eqref{power_problem} is reformulated as $a_k p_k^2 + b_k p_k + c_k \le 0$. By setting $k = k_{\rm max}$ and $ p_k = p_{\rm max} $, we have $a_{k_{\rm max}} = \varepsilon_{\rm eff}^2 \beta_{k_{\rm max}}^2 \sigma_{k_{\rm max}}^2 - \rho_{k_{\rm max}} \sigma_{k_{\rm max}}^4 $, $b_{k_{\rm max}} =  \sigma_{k_{\rm max}}^2 + 2 \varepsilon_{\rm eff} \beta_{k_{\rm max}} \sigma_{k_{\rm max}} - 2 \rho_{k_{\rm max}} \sigma^2_{k_{\rm max}}$, and $c_k = 1- \rho_k$. Having obtained $p_{\rm max}$, we have $a_{k} = - \rho_{k} \sigma_{k}^4 $ and $b_{k} =  \sigma_{k}^2 + 2 \varepsilon_{\rm eff} \beta_{k} \sigma_{k} + \varepsilon_{\rm eff}^2 \beta_{k}^2 \sigma_{k}^2 p_{\rm max} - 2 \rho_{k} \sigma^2_{k}$, for $k \neq k_{\rm max}$. The feasibility condition gives $a_{k_{\rm max}}\le 0$, and we also have $c_k\ge 0$. Hence, the quadratic function includes one positive and one negative root. {The positive root is}
\begin{equation}\label{power_root}
p_k^{\star} = \frac{1}{2a_k}\left( -b_k -\sqrt{b_k^2 - 4 a_k c_k} \right), \quad \forall k.
\end{equation}

\subsection{Summary}
Our low-complexity HB scheme is {outlined} in Algorithm \ref{less_complex}. The number of data streams is initially set to $N^{\rm RF}_{\rm t}$, and {the} RF beamformer matrices are obtained as described in Section \ref{analog_section}. Next, the effective channel is computed, the infeasible data stream with {the} highest violation is found and discarded, and {the} RF beamforming is re-done when $N_{\rm s}\ge\frac{N^{\rm RF}_{\rm t}}{2}$. This procedure is repeated until all remaining data streams are feasible. Next, the data stream with highest transmit power in {the scenario with perfect CSI} is identified and its transmit power in {the scenario with imperfect CSI} is computed from \eqref{power_root}, {after which we calculate the} transmit power levels of other data streams. Finally, we obtain {the} baseband precoder as ${\bf F}_{\rm BB}^{(1)} = \bf V P^{\frac{1}{2}}$ and {the} baseband combiner from \eqref{bb_combining}.

\begin{algorithm}[t!]
\begin{small}
	\label{less_complex}
	\DontPrintSemicolon
	\KwInput
	{
		Random ${\bf{F}}_{\rm{BB}}^{(2)}$ and ${\bf{W}}_{\rm{BB}}^{(2)}, {\widehat{\bf H}}, \varepsilon_{\rm eff}, N^{\rm RF}_{\rm t}, {\boldsymbol\rho}$
	}
	$N_{\rm s} = N^{\rm RF}_{\rm t}$\\
	Obtain ${\bf{F}}_{\rm{RF}}{\bf{F}}_{\rm{BB}}^{(2)}$ and ${\bf{W}}_{\rm{RF}}{\bf{W}}_{\rm{BB}}^{(2)}$ as described in Section \ref{analog_section}\\
	Compute $\widehat{\bf H}_{\rm eff} = {\bf{W}}_{\rm{BB}}^{\rm (2)H} {\bf{W}}_{\rm{RF}}^{\rm H} \widehat{\bf H}{\bf{F}}_{\rm{RF}}{\bf{F}}_{\rm{BB}}^{(2)}$\\
	Compute SVD of ${\bf R}_{\rm n}^{-\frac{1}{2}}\widehat{\bf H}_{\rm eff}$ and return $\bf V$ and $\boldsymbol \Sigma$\\
	\If {$\max\limits_k \frac{1}{\sigma^2_k}\varepsilon_{\rm eff}^2\beta_k^2 - \rho_k >0$ }{
	$\tilde{ k} = \argmax\limits_k \frac{1}{\sigma^2_k}\varepsilon_{\rm eff}^2\beta_k^2  - \rho_k$\\
	Discard  $\tilde{ k}^{\rm th}$ data stream from the set of data streams\\
	$ N_{\rm s} = N_{\rm s} -1$ \\
	\If {$N_{\rm s} > \frac{N^{\rm RF}_{\rm t}}{2}$}{
		return to \textbf{2}\\
		}
	}
	Obtain $k_{\rm max} = \argmax\limits_{k} \frac{1}{\sigma_k^2}\left( \frac{1}{\rho_k}-1 \right)$\\
	Compute $a_{k_{\rm max}}, b_{k_{\rm max}}, c_{k_{\rm max}}, \text{ and } p_{\rm max}$ in \eqref{power_root}\\
		\For{$k=1$ \rm{to} $N_{\rm s} \wedge k \neq k_{\rm max}$}
		{
		Compute $a_k, b_k, c_k, \text{ and } p_k$ in \eqref{power_root}\\
		}
	Compute ${\bf F}_{\rm BB}^{(1)} = \bf V P^{\frac{1}{2}}$\\
	Compute ${\bf W}_{\rm BB}^{(1)}$ from \eqref{bb_combining}\\
	\KwOutput{$ {\bf F}_{\rm BB}^{(1)}, {\bf{F}}_{\rm{BB}}^{(2)}, {\bf{F}}_{\rm{RF}}, {\bf W}_{\rm BB}^{(1)},{\bf{W}}_{\rm{BB}}^{(2)}, \text{ and }{\bf{W}}_{\rm{RF}} $}
	\caption{Low-Complexity Scheme for Solving \eqref{qos_problem}}
\end{small}
\end{algorithm}
\vspace{-5 pt}
\section{Multi-User Hybrid Beamforming}\label{mu_section}

In this section, we extend our iterative and low-complexity { HB schemes} to multi-user systems.
\vspace{-10 pt}
\subsection{System Model}

The multi-user system includes one transmitter equipped with $N_{\rm t}$ antennas and $N_{\rm t}^{\rm RF}$ RF chains, and $U$ receivers each with $N_{\rm r}$ antennas and $N_{\rm r}^{\rm RF}$ RF chains. The vector ${{\bf s}=[{\bf s}_1^{\rm T}, {\bf s}_2^{\rm T},\ldots, {\bf s}_U^{\rm T}]^{\rm T}}$ is transmitted, where ${\bf s}_u\in \mathbb{C}^{N_{\rm{s}} \times 1}$. The estimated vector of {the} received symbols at any given instance by receiver $u$ is
\begin{equation}\label{s_mu}
\hat{\bf{s}}_u = {\bf{W}}_{{\rm BB},u}^{\rm H}{\bf{W}}_{{\rm RF},u}^{\rm H}{\bf H}_u{ \bf{F}}_{\rm{RF}}{\bf F}_{\rm BB}{\bf{s}} + {\bf{W}}_{{\rm BB},u}^{\rm H}{\bf{W}}_{{\rm RF},u}^{\rm H}{\bf n}_u,
\end{equation}
where ${\bf{W}}_{{\rm BB},u}$ and ${\bf{W}}_{{\rm RF},u}$ are the baseband combiner and phase {shifter} matrices of receiver $u$, respectively; ${\bf H}_u$ is the channel gain matrix between the transmitter and receiver $u$, and ${\bf n}_u$ is the noise vector at receiver $u$. CSI is assumed to be erroneous, i.e., ${\bf H}_u = \widehat{\bf H}_u + \boldsymbol \Delta_u$, and ${\rm Pr}\{\|\boldsymbol\Delta_u \|_{\rm F}\le \varepsilon_u\} =  P_{\rm in}$.

The baseband precoder and combiner matrices are divided into two stages each, {such that} ${\bf F}_{\rm BB} = {\bf F}_{\rm BB}^{(2)}{\bf F}_{\rm BB}^{(1)}$ and ${\bf W}_{{\rm BB},u} = {\bf W}_{{\rm BB},u}^{(2)}{\bf W}_{{\rm BB},u}^{(1)}$, where ${\bf{F}}_{\rm{BB}}^{(1)}=\left[{\bf{F}}_{\rm{BB},1}^{(1)},\ldots, {\bf{F}}_{\rm{BB},U}^{(1)}\right]$, and ${\bf{F}}_{{\rm BB},u}^{(1)}\in \mathbb{C}^{UN_{\rm{s}} \times N_{\rm{s}}}$ is the baseband precoder for receiver $u$. We wish to minimize the transmit power while satisfying MSE constraints {(i.e., $[{\bf E}_u]_{k,k}\le \rho_{u,k}$)}, where ${\bf E}_u$ is {the} MSE matrix estimated by receiver $u$.
\vspace{-3 mm}
\subsection{RF Beamforming}

{The} RF combiner of receiver $u$ is obtained by approximating the left eigenvectors of $\widehat{\bf H}_u$ derived by SVD $\widehat{\bf H}_u = \widehat{\bf U}_u \widehat{\boldsymbol \Sigma}_u \widehat{\bf V}_u^{\rm H}$. In other words, we use RF beamforming in Section \ref{analog_section} with proper substitutions to find ${\bf W}_{{\rm RF},u}$ and ${\bf W}_{{\rm BB},u}^{\rm (2)}$ by minimizing $\|{\bf W}_{{\rm RF},u} {\bf W}_{{\rm BB},u}^{\rm (2)} - \widehat{\bf U}_u \|^2_{\rm F}$. {The} RF precoder is then obtained by minimizing $\|{\bf F}_{{\rm RF}} {\bf F}_{{\rm BB}}^{\rm (2)} - {\bf V}_{\rm mu} \|^2_{\rm F}$, where ${\bf V}_{\rm mu}$ includes the eigenvectors of $\widehat{\bf H}_{\rm mu}^{\rm H}{\bf W}_{\rm mu}{\bf W}_{\rm mu}^{\rm H}\widehat{\bf H}_{\rm mu}$ that correspond to {the} $U \times N_{\rm s}$ largest eigenvalues, $\widehat{\bf H}_{\rm mu}$ is the estimated multi-user channel $\widehat{\bf H}_{\rm mu} = [\widehat{\bf H}_1^{\rm T},\ldots, \widehat{\bf H}_U^{\rm T}]^{\rm T}$, and ${\bf W}_{\rm mu} = {\rm diag}({\bf W}_{{\rm RF},1} {\bf W}_{{\rm BB},1}^{\rm (2)}, \ldots, {\bf W}_{{\rm RF},U} {\bf W}_{{\rm BB},U}^{\rm (2)})$.

The effective channel of receiver $u$ is ${\bf H}_{{\rm eff},u} = \widehat{\bf H}_{{\rm eff},u} + {\boldsymbol \Delta}_{{\rm eff},u}$, where $\widehat{\bf H}_{{\rm eff},u} = {\bf W}_{{\rm BB},u}^{\rm (2)H}{\bf W}_{{\rm RF},u}^{\rm H}\widehat{\bf H}_u\times$
$ {\bf F}_{{\rm RF}} {\bf F}_{{\rm BB}}^{\rm (2)}$ and ${\boldsymbol \Delta}_{{\rm eff},u} = {\bf W}_{{\rm BB},u}^{\rm (2)H}{\bf W}_{{\rm RF},u}^{\rm H}\boldsymbol \Delta_u{\bf F}_{{\rm RF}} {\bf F}_{{\rm BB}}^{\rm (2)}$. Similar to \eqref{effective_norm}, we have\vspace{-5 pt}
\begin{multline}\label{delta_eq}
\|\boldsymbol\Delta_{{\rm eff},u}\|_{\rm F}^2 = \boldsymbol\delta_u^{\rm H} {\bf A}_{{\rm RF},u} \boldsymbol\delta_u =\\
\boldsymbol\delta_u^{\rm H} {\bf U}_{{\rm RF},u} {\boldsymbol \Lambda}_{{\rm RF},u} {\bf U}_{{\rm RF},u}^{\rm H} \boldsymbol\delta_u = \tilde{\boldsymbol\delta}_u^{\rm H} {\boldsymbol \Lambda}_{{\rm RF},u}  \tilde{\boldsymbol\delta}_u,
\end{multline}
where $\boldsymbol\delta_u = {\rm vec}(\boldsymbol \Delta_u)$, ${\bf A}_{{\rm RF},u} = {\bf A}_{\rm F} \otimes {\bf A}_{{\rm W},u}$, and ${\bf A}_{{\rm W},u} = {\bf W}_{{\rm RF},u}{\bf W}_{{\rm BB},u}^{(2)}{\bf W}_{{\rm BB},u}^{\rm (2) H}{\bf W}_{{\rm RF},u}^{\rm H}$. Therefore, $\|\boldsymbol\Delta_{{\rm eff},u}\|_{\rm F}^2 = \sum_{n=1}^{UN_{\rm s}^2}\lambda^u_{{\rm RF},n}|\tilde{\delta}_n^u|^2 \approx \sum_{n=1}^{UN_{\rm s}^2}|\tilde{\delta}_n^u|^2$, where $\tilde{\delta}_n^u$ is the entry $n$ of $\boldsymbol\delta_u$ and $\lambda^u_{{\rm RF},n}$ is the $n^{\rm th}$ eigenvalue of ${\bf A}_{{\rm RF},u}$. As a result, $\|\boldsymbol\Delta_{{\rm eff},u}\|_{\rm F}^2$ has Erlang distribution with shape $U \times N^2_{\rm s}$, rate $\sigma_e^{-2}$, and CDF $\phi_3(\cdot)$. Hence, ${\rm Pr}\{\|\boldsymbol\Delta_{{\rm eff},u} \|_{\rm F}\le \varepsilon_{{\rm eff}}\} =  P_{\rm in}$, which implies $\varepsilon_{\rm eff}^2 = \phi_3^{-1}(P_{\rm in})$.
\vspace{-10 pt}
\subsection{Multi-User Iterative HB Scheme}
We extend our iterative scheme in Section \ref{iterative_section} to {multi-user (MU)-MIMO} systems. From \eqref{s_mu}, we have
\begin{multline}\label{s_eff}
\hat{\bf{s}}_u = {\bf{W}}_{{\rm BB},u}^{\rm (1)H}{\bf H}_{{\rm eff}, u}{\bf F}_{{\rm BB},u}^{(1)}{\bf s}_u +\\
 \sum \limits_{u' \neq u} {\bf{W}}_{{\rm BB},u}^{\rm H(1)}{\bf H}_{{\rm eff}, u}{\bf F}_{{\rm BB},u'}^{(1)}{\bf s}_{u'} + {\bf{W}}_{{\rm BB},u}^{\rm (1)H}\widetilde{\bf n}_u,
\end{multline}
where $\widetilde{\bf n}_u = {\bf W}_{{\rm BB},u}^{\rm (2) H}{\bf W}_{{\rm RF},u}^{\rm H}{\bf n}_u$. After some manipulations, we get
\begin{equation}\label{e_mu}
\begin{array}{ll}
{\bf E}_u =& {\mathbb E} \left\{({\bf s}_u - \hat{\bf s}_u)({\bf s}_u - \hat{\bf s}_u)^{\rm H}  \right\} = \\
&{\bf W}_{{\rm BB},u}^{\rm (1)H}{\bf H}_{{\rm eff}, u}{\bf F}_{\rm BB}^{(1)}{\bf F}_{\rm BB}^{\rm (1)H}{\bf H}_{{\rm eff}, u}^{\rm H} {\bf W}_{{\rm BB},u}^{\rm (1)} -\\
& 2 {\rm Re}\left\{ {\bf W}_{{\rm BB},u}^{\rm (1)H}{\bf H}_{{\rm eff}, u}{\bf F}_{{\rm BB},u}^{(1)} \right\} +\\
&{\bf{W}}_{{\rm BB},u}^{\rm (1)H}{\bf R}_{{\rm n},u}{\bf{W}}_{{\rm BB},u}^{\rm (1)} + {\bf I}_{N_{\rm s}},
\end{array}
\end{equation}
where ${\bf R}_{{\rm n},u} = \sigma_{\rm n}^2{\bf W}_{{\rm BB},u}^{\rm (2) H}{\bf W}_{{\rm RF},u}^{\rm H} {\bf W}_{{\rm BB},u}^{\rm (2)}{\bf W}_{{\rm RF},u}$.

In what follows, we analyze the feasibility of our MU-MIMO HB. Similar to the  single-user MIMO HB, we introduce a set of non-negative variables, and solve the sample problem \eqref{sampled_problem_mu},
\begin{figure*}
\begin{equation}\label{sampled_problem_mu}
\begin{array}{ccc}
& \mathop{\min}\limits_{{\bf F}_{\rm BB}^{\rm (1)},\{{\bf W}_{{\rm BB},u}^{\rm (1)}\}_{u=1}^U,{\boldsymbol \alpha}} &  \gamma{\rm Tr}\left(  {\bf{F}}_{\rm{BB}}^{\rm (1)H}{\bf A}_{\rm P}{\bf{F}}_{\rm BB}^{(1)} \right) + (1-\gamma)\sum\limits_{u=1}^U\sum\limits_{k=1}^{N_{\rm s}}\alpha_{u,k}\\
& \text{subject to}&\left\{
\begin{array}{l}
{\bf E}_{u,k}({\bf F}_{\rm BB}^{\rm (1)},{\bf W}_{{\rm BB},u}^{\rm (1)};{\boldsymbol \Delta}_{u,k}^{\rm wc})\le \rho_{u,k} + \alpha_{u,k}, \quad \forall u, k \text{ and } \forall {\boldsymbol \Delta}_{u,k}^{\rm wc}\in \mathcal{D}_{u,k}^q\\
\alpha_{u,k}\ge 0, \quad \forall u,k,
\end{array}
\right.
\end{array}
\end{equation}
\hrulefill
\end{figure*}
where ${\bf E}_{u,k}(\cdot)$ is {the} MSE of data stream $k$ estimated by receiver $u$. Solving \eqref{sampled_problem_mu} is similar to solving \eqref{sampled_problem} via Algorithm \ref{sampled_algorithm}. However, we need to find quadratic expressions for ${\bf E}_{u,k}(\cdot)$ in terms of ${\bf F}_{\rm BB}^{\rm (1)}$ (similar to \eqref{qcqp_problem}) and ${\bf W}_{{\rm BB},u}^{\rm (1)}$ (similar to \eqref{bbc_problem}). From \eqref{e_mu}, we get
\begin{equation}\label{bbp_mu}
{\bf E}_{u,k} = {\bf f}_{\rm BB}^{\rm (1)H}{\bf A}_{u,k}^{\rm W} {\bf f}_{\rm BB}^{(1)}-2{\rm Re}\left\{ {\bf f}_{u,k}^{\rm (1)H}{\bf b}_{u,k}^{\rm W} \right\}+{\bf{w}}_{u,k}^{\rm (1)H}{\bf R}_{{\rm n},u}{\bf{w}}_{u,k}^{(1)}+1,
\end{equation}
where ${\bf f}_{\rm BB}^{(1)} = {\rm vec}({\bf F}_{\rm BB}^{(1)})$, ${\bf f}_{u,k}^{\rm (1)}$ is the $((u-1)N_{\rm s}+k)^{\rm th}$ column of ${\bf F}_{\rm BB}^{(1)}$, ${\bf w}_{u,k}^{(1)}$ is column $k$ of ${\bf W}_{{\rm BB},u}^{\rm (1)}$, ${\bf A}_{u,k}^{\rm W} = {\bf I}_{UN_{\rm s}} \otimes {\bf H}_{u,k}^{\rm wcH}{\bf{w}}_{u,k}^{(1)}{\bf{w}}_{u,k}^{\rm (1)H}{\bf H}_{u,k}^{\rm wc}$, ${\bf b}_{u,k}^{\rm W} = {\bf H}_{u,k}^{\rm wcH}{\bf{w}}_{u,k}^{(1)}$, and ${\bf H}_{u,k}^{\rm wc} = \widehat{\bf H}_{{\rm eff}, u}+ {\boldsymbol \Delta}_{u,k}^{\rm wc}$. The quadratic expression for ${\bf E}_{u,k}(\cdot)$ in terms of ${\bf W}_{{\rm BB},u}^{\rm (1)}$ is\vspace{-5 pt}
\begin{equation}\label{bbc_mu}
{\bf E}_{u,k} = {\bf w}_{u,k}^{\rm (1)H}{\bf A}_{u}^{\rm F} {\bf w}_{u,k}^{(1)}-2{\rm Re}\left\{ {\bf w}_{u,k}^{\rm (1)H}{\bf b}_{u,k}^{\rm F} \right\}+1,
\end{equation}
where ${\bf A}_{u}^{\rm F} = {\bf H}_{u,k}^{\rm wc}{\bf F}_{\rm BB}^{(1)}{\bf F}_{\rm BB}^{\rm (1)H} {\bf H}_{u,k}^{\rm wcH}+{\bf R}_{{\rm n},u}$ and ${\bf b}_{u,k}^{\rm F} = {\bf H}_{u,k}^{\rm wc}{\bf f}_{u,k}^{\rm (1)}$.

{The worst-case analysis} requires finding ${\boldsymbol \Delta}_{u,k}^{\rm wc}$. Similar to \eqref{mse_delta}, we write
\begin{equation}\label{mse_delta_mu}
{\bf E}_{u,k} = {\boldsymbol \delta}_{{\rm eff},u}^{\rm H} {\bf A}^{\delta}_{u,k} {\boldsymbol \delta}_{{\rm eff},u} +  2{\rm Re}\{ {\boldsymbol \delta}_{{\rm eff},u}^{\rm H}{\bf b}^{\delta}_{u,k}\} + \widehat{\bf E}_{u,k},
\end{equation}
\begin{figure*}
\begin{equation}\label{mse_bb_mu}
\begin{array}{ll}
{\bf E}_{u}= &\widehat{\bf E}_{u} +
2\underbrace{{\rm Re}\left\{{\bf W}_{{\rm BB},u}^{\rm (1)H}{\boldsymbol \Delta}_{{\rm eff},u}\left[\bar{\bf F}_u\widehat{\bf H}_{{\rm eff}, u}^{\rm H}{\bf W}_{{\rm BB},u}^{\rm (1)} + {\bf F}_{{\rm BB},u}^{\rm (1)}\left({\bf F}_{{\rm BB},u}^{\rm (1)H}\widehat{\bf H}_{{\rm eff},u}^{\rm H}{\bf W}_{{\rm BB}, u}^{\rm (1)}-{\bf I}_{N_{\rm s}}\right)\right]\right\}}_{{\bf X}_u}+\\
&\underbrace{{\bf W}_{{\rm BB}, u}^{\rm (1)H}{\boldsymbol \Delta}_{{\rm eff},u}{\bf F}_{\rm BB}^{(1)}{\bf F}_{\rm BB}^{\rm (1)H}{\boldsymbol \Delta}_{{\rm eff},u}^{\rm H}{\bf W}_{{\rm BB}, u}^{\rm (1)}}_{{\bf Y}_u},
\end{array}
\end{equation}
\hrulefill
\end{figure*}
where ${\boldsymbol \delta}_{{\rm eff},u} = {\rm vec}({\boldsymbol \Delta}_{{\rm eff},u})$, ${\bf A}^{\delta}_{u,k} = {\bf F}_{\rm BB}^{(1)*}{\bf F}_{\rm BB}^{\rm (1)T} \otimes {\bf w}_{u,k}^{\rm (1)}{\bf w}_{u,k}^{\rm (1)H}$, ${\bf b}^{\delta}_{u,k} = ({\bf F}_{\rm BB}^{(1)}{\bf F}_{\rm BB}^{\rm (1)H} \widehat{\bf H}_{{\rm eff},u}^{\rm H}{\bf w}_{u,k}^{\rm (1)} - {\bf f}_{u,k}^{\rm (1)})^* \otimes {\bf w}_{u,k}^{\rm (1)}$, and $\widehat{\bf E}_{u,k}$ is obtained from \eqref{e_mu} by replacing ${\bf H}_{{\rm eff}, u}$ with $\widehat{\bf H}_{{\rm eff}, u}$. Finding the maximizer of \eqref{mse_delta_mu}, i.e., ${\boldsymbol \Delta}_{u,k}^{\rm wc}$, is similar to finding the worst-case effective channel for the single-user HB.

All {of the} steps in our iterative scheme for single-user HB in Fig. \ref{flowchart2} are applicable to multi-user HB with appropriate substitutions as described above. As such, extending our discussions on convergence and CC of our iterative scheme from single-user HB to multi-user HB is straightforward.\vspace{-3 mm}
\subsection{Low-Complexity HB for Multi-User MIMO Systems}

Similar to Section \ref{low_section}, we introduce ${\bf F}_{{\rm BB},u}^{\rm (1)}$ and ${\bf W}_{{\rm BB},u}^{\rm (1)}$, and derive the sufficient condition for feasibility of each individual data stream. We adopt block diagonalization (BD) for {the} baseband precoder, i.e., ${\bf F}_{{\rm BB},u}^{\rm (1)} = \bar{\bf V}_u^{(0)}{\bf V}_u {\bf P}_{u}^{\frac{1}{2}}$, where the columns of $\bar{\bf V}_u^{(0)}$ span the null space of $\bar{\bf H}_u = [\widehat{\bf H}_{{\rm eff}, 1},\ldots, \widehat{\bf H}_{{\rm eff}, u-1}, \widehat{\bf H}_{{\rm eff}, u+1},\ldots,\widehat{\bf H}_{{\rm eff}, U}]$, obtained by SVD of $\bar{\bf H}_u = \bar{\bf U}_u [\bar{\boldsymbol \Sigma}_u  \quad {\bf 0}][\bar{\bf V}_u^{(1)} \quad \bar{\bf V}_u^{(0)}]^{\rm H}$. The value of ${\bf V}_u$ is obtained by SVD of  ${\bf R}_{{\rm n},u}^{-\frac{1}{2}}\widehat{\bf H}_{{\rm eff}, u}\bar{\bf V}_u^{(0)} = {\bf U}_u {\boldsymbol \Sigma}_u {\bf V}_u$.

For MMSE combining, i.e., ${\bf W}_{{\rm BB},u}^{(1)} = \left( \widehat{\bf H}_{{\rm eff},u}{\bf F}_{{\rm BB},u}^{(1)}{\bf F}_{{\rm BB},u}^{\rm (1)H} \widehat{\bf H}_{{\rm eff},u}^{\rm H} + {\bf R}_{{\rm n},u}\right)^{-1} \widehat{\bf H}_{{\rm eff},u}{\bf F}_{{\rm BB},u}^{(1)}$ in the baseband combiner of receiver $u$, and from \eqref{e_mu}, we get \eqref{mse_bb_mu}, where $\bar{\bf F}_u = \sum \limits_{u' \neq u} {\bf F}_{{\rm BB},u'}^{\rm (1)}{\bf F}_{{\rm BB},u'}^{\rm (1)H}$. By BD precoding, we have $\widehat{\bf H}_{{\rm eff}, u}{\bf F}_{{\rm BB},u'}^{\rm (1)} = \bf 0$, $\forall u'\neq u$, resulting in $\bar{\bf F}_u\widehat{\bf H}_{{\rm eff}, u}^{\rm H} = \bf 0$. Hence,
\begin{equation}
{\bf X}_u = {\rm Re}\left\{{\bf W}_{{\rm BB},u}^{\rm (1)H}{\boldsymbol \Delta}_{{\rm eff},u}{\bf F}_{{\rm BB},u}^{\rm (1)}\left({\bf F}_{{\rm BB},u}^{\rm (1)H}\widehat{\bf H}_{{\rm eff},u}^{\rm H}{\bf W}_{{\rm BB}, u}^{\rm (1)}-{\bf I}_{N_{\rm s}}\right)\right\},
\end{equation}
which is similar to \eqref{mse_bb}. Therefore, the maximum value of $[{\bf X}_u]_{k,k}$ is
\begin{equation}\label{x_wc_mu}
[{\bf X}^{\rm wc}_u]_{k,k} = \frac{\sigma_{u,k}  p_{u,k}}{(1+\sigma_{u,k}^2  p_{u,k})^2} \beta_{u,k} \varepsilon_{\rm eff},
\end{equation}
where  $\beta_{u,k} =\sqrt{{\bf u}^{\rm H}_{u,k} {\bf R}_{{\rm n},u}^{-1} {\bf u}_{u,k}}$, ${\bf u}_{u,k}$ is column $k$ of ${\bf U}_u$, and $\sigma_{u,k}$ and $p_{u,k}$ are the $k^{\rm th}$ diagonal elements of ${\boldsymbol \Sigma}_u$ and ${\bf P}_u$, respectively.

BD baseband precoding gives ${\bf F}_{\rm BB}^{(1)}{\bf F}_{\rm BB}^{\rm (1)H} = \sum_{u=1}^U  \bar{\bf V}_u^{(0)}{\bf V}_u {\bf P}_{u}{\bf V}_u^{\rm H} \bar{\bf V}_u^{\rm (0)H} = {\bf V}_{\rm mu} {\bf P}_{\rm mu} {\bf V}_{\rm mu}^{\rm H}$, where ${\bf V}_{\rm mu} = [\bar{\bf V}_1^{(0)}{\bf V}_1,\ldots,\bar{\bf V}_U^{(0)}{\bf V}_U]$ and ${\bf P}_{\rm mu} = {\rm diag}({\bf P}_{1},\ldots,{\bf P}_{U})$. Hence,\vspace{-5 pt}
\begin{equation}
{\bf Y}_u = {\bf W}_{{\rm BB}, u}^{\rm (1)H}{\boldsymbol \Delta}_{{\rm eff},u}{\bf V}_{\rm mu} {\bf P}_{\rm mu} {\bf V}_{\rm mu}^{\rm H}{\boldsymbol \Delta}_{{\rm eff},u}^{\rm H}{\bf W}_{{\rm BB}, u}^{\rm (1)},
\end{equation}
which is similar to \eqref{mse_bb}. Therefore, the maximum value of $[{\bf Y}_u]_{k,k}$ is\vspace{-5 pt}
\begin{equation}\label{y_wc_mu}
[{\bf Y}_u^{\rm wc}]_{k,k} = \frac{\sigma_{u,k}^2  p_{u,k} p_{\rm max}}{(1+\sigma_{u,k}^2  p_{u,k})^2}\varepsilon_{\rm eff}^2\beta_{u,k}^2,
\end{equation}
where $p_{\rm max} = \max\limits_{u,k} p_{u,k}$. {With} BD baseband precoding and MMSE baseband combining, we have $\widehat{\bf E}_{u,k} = \frac{1}{1+\sigma_{u,k}^2 p_{u,k}}$. {The worst-case MSE} in \eqref{mse_bb_mu} is similar to {the} worst-case MSE in \eqref{wc_mse} by substituting subscript $k$ with $u,k$. As a result, {the} feasibility analysis and transmit power allocation in our low-complexity scheme in Section \ref{low_section} are also valid for MU-MIMO HB.
\vspace{-3 mm}
\section{Simulations}\label{simulation_section}

{For the following simulations, we consider the extended Saleh-Valenzuela channel model} \cite{tajalli2021qos} {with three paths, each with ten rays and normal distributed complex gain and Laplacian distributed angles of arrival and departure with ten degree spreads. We also assume uniform linear arrays with half-wavelength inter-element spacing. The covariance matrix of the noise vector is identity; that is, $\sigma_{\rm n} = 1$. We assume the same value of the highest acceptable MSE for all data streams; that is, $\rho = \rho_k , \forall k$.}

\begin{table*}
	\centering
	\caption{Size of uncertainty region for {antennas with different number of elements}}
	\label{set_size}
	\begin{tabular}{ c c c c c c c}	
		\hline
		$N_{\rm t} \times N_{\rm r}$ & $64 \times 36$ & $100 \times 36$ & $144 \times 36$ & $100 \times 64$ & $144 \times 64$ & $256 \times 64$\\\hline		
		$\varepsilon$ & $4.97$ & $6.15$ &  $7.35$ & $8.15$ & $9.75$ & $12.90$\\\hline
	\end{tabular}
\end{table*}

\begin{figure}[!t]
	\centering
	\includegraphics[width=7 cm,height=5.5 cm]{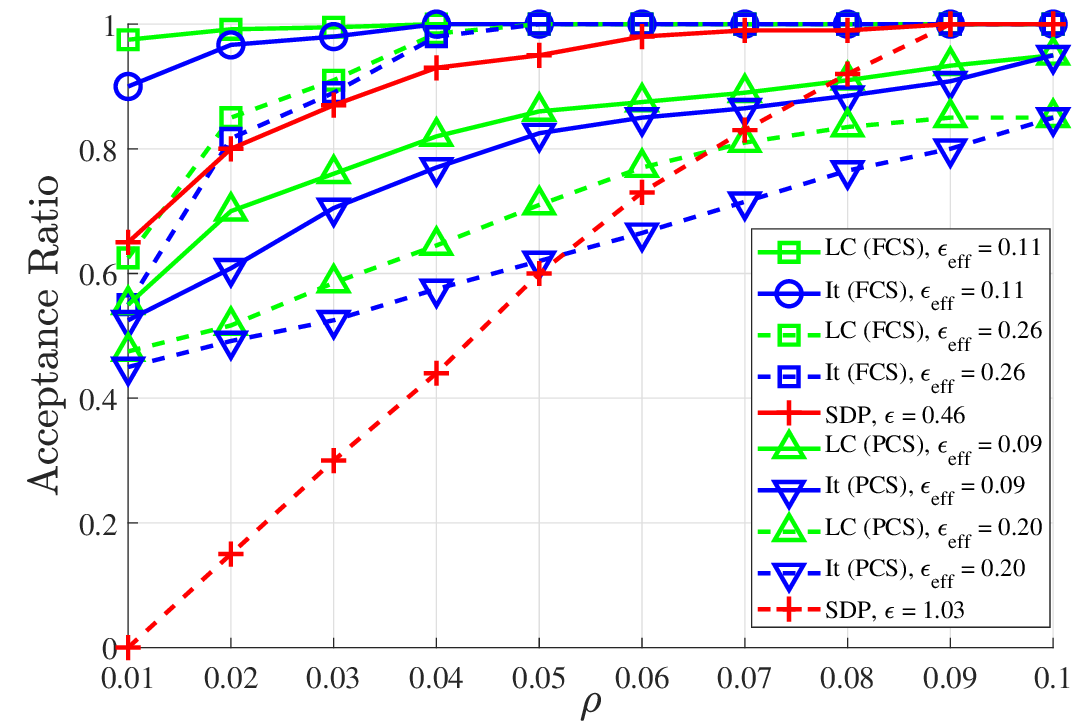}
	\caption{{Acceptance ratio of our iterative and low complexity HB methods.}}
	\label{ar_su}
\end{figure}

{We compare the size of the effective channel's uncertainty region with that of the channel itself for antennas with different number of elements. In doing so, we consider a single-user system with $N_{\rm s}=N_{\rm RF}=6$, where $N_{\rm RF} = N^{\rm RF}_{\rm t} = N^{\rm RF}_{\rm r}$, for which when $P_{\rm in} = 1-10^{-3}$ and $\sigma_{\rm e}^2 = 10^{-2}$, we find the values of $\varepsilon_{\rm eff}$ and $\varepsilon$ such that ${\rm Pr}\{\|\bf{\Delta}\|_{\rm F} \le \varepsilon \}  = {\rm Pr}\{\|\bf{\Delta}_{\rm eff}\|_{\rm F} \le \varepsilon_{\rm eff} \}=P_{\rm in}$. The size of uncertainty region for different systems is shown in Table} \ref{set_size}{. For each entry, we first find the RF precoder and combiner for the given channel, and then obtain CDFs of $\|\bf{\Delta}\|_{\rm F}$ and $\|\bf{\Delta}_{\rm eff}\|_{\rm F}$ for $10^4$ channel realizations. Next, we find the uncertainty region for when $P_{\rm in} = 1-10^{-3}$. When antenna arrays have more elements, the uncertainty region is expanded for a given $P_{\rm in}$. However, from }\eqref{delta_sum}{, the CDF of $\|\bf{\Delta}_{\rm eff}\|_{\rm F}$ depends only on $N_{\rm s}$, and thus does not vary with the number of antenna elements. For all antenna arrays in Table} \ref{set_size}{, we have $\varepsilon_{\rm eff} = 0.72$ in the FCS. More transmit power is required to overcome higher  uncertainties} \cite{vucic2009robust}{, and using $\varepsilon_{\rm eff}$ instead of $\varepsilon$ significantly reduces the transmit power when the number of antenna elements is high.}

We now compare the performance of our {schemes} with that of {the} fully digital robust beamforming in \cite{vucic2009robust}, where in the latter, uncertainty is formulated by way of a linear matrix inequality (LMI), and the problem is solved via SDP. {The} CC of \cite{vucic2009robust} is very high when the size of matrices in LMIs are large, which is the case in fully digital massive MIMO beamforming. In {our} simulations, we assume that the transmitter and receiver have $20$ and $8$ antenna elements, respectively, and $N_{\rm s} = 2$.

{The acceptance ratio (AR) is the ratio of the number of feasible data streams to all data streams. We obtain $\varepsilon$ in the SDP-based scheme and $\varepsilon_{\rm eff}$ in our scheme for a given $P_{\rm in}$ and $\sigma_{\rm e}$, and compute the corresponding values of AR, which are given in Fig.} \ref{ar_su} {for different values of the highest acceptable MSE $\rho$. In Fig.} \ref{ar_su}, {for $P_{\rm in} = 1-10^{-4}$ and $\sigma_{\rm e}^2 = 10^{-3}, 5\times 10^{-3}$, we have $\varepsilon = 0.463, 1.034, \varepsilon_{\rm eff} = 0.11, 0.26$ in the FCS, and $\varepsilon_{\rm eff} = 0.08, 0.20$ in the PCS, respectively.} Fig. \ref{ar_su} shows that {the} AR is smaller for larger uncertainty regions in all schemes because {the} worst-case MSE is higher when it is maximized over a larger uncertainty region, resulting in more infeasible data streams. {As we can see,} our HB schemes outperform the SDP-based fully digital beamforming {scheme} in \cite{vucic2009robust} {that} is insensitive to {the} individual MSE constraints, which means that {by using the scheme in} \cite{vucic2009robust}, the problem is either feasible or infeasible for all data streams in each channel realization. {By contrast, in our schemes, the feasibility of each MSE constraint is considered separately, resulting in a higher AR. This type of feasibility analysis is quite novel; it has not been proposed in the existing literature.} Our low-complexity scheme achieves a slightly better AR than our iterative scheme because the objective function in \eqref{feasible_problem} sacrifices the AR in favor of transmit power. {Our schemes achieve a higher AR in the FCS than in the PCS because channel eigenvalues are better approximated and the effective channel gain is higher.}

\begin{figure}[!t]
	\centering
	\includegraphics[width=7 cm,height=5.5 cm]{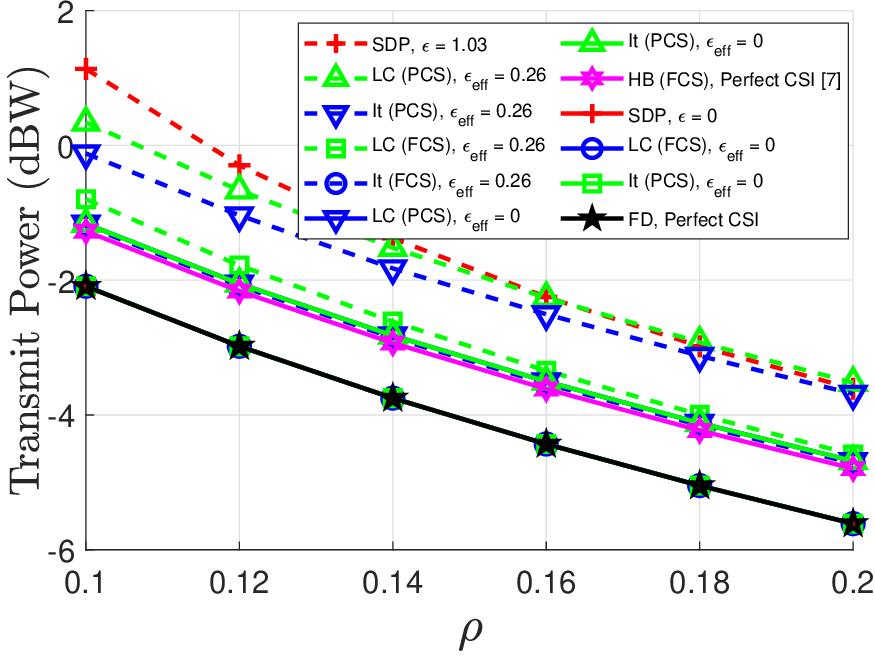}
	\caption{{Transmit power of our iterative and low complexity HB methods. }}
	\label{txpower_su}
\end{figure}

Fig. \ref{txpower_su} {presents a comparison of the transmit power in the SDP-based scheme in} \cite{vucic2009robust} relative to our schemes for both perfect and imperfect CSI scenarios. The comparison is fair when the number of data streams in different schemes is the same. Hence, we only consider channel realizations for which $\rm{AR}=1$. In the FCS with perfect CSI ($\varepsilon=\varepsilon_{\rm eff}=0$), our schemes consume the same amount of power as in the SDP-based and the optimal fully digital schemes, but less than the power consumed in \cite{wang2022joint}. With imperfect CSI, however, our schemes consume less power than the SDP-based scheme. The transmit power in our low-complexity scheme is slightly higher relative to our iterative scheme because in the former, we maximize $\bf{X}$ and $\bf{Y}$ in \eqref{mse_bb} separately, {whereas the latter is based on} \eqref{wc_error}. In other words, our low-complexity scheme requires more transmit power to overcome higher MSE values. 

\begin{figure}[t]
	\centering
	\includegraphics[width=7 cm,height=5.5 cm]{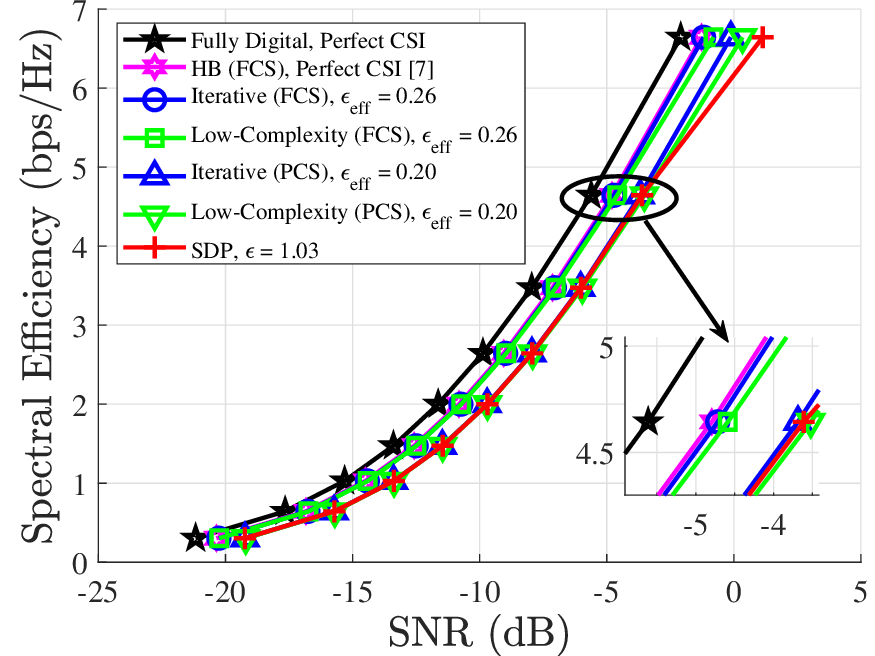}
	\caption{{Worst-case spectral efficiency of our HB schemes for $N_{\rm s}=2$ and $N_{\rm RF}=4$.}}
	\label{rate_snr}
\end{figure}

Fig. \ref{rate_snr} {offers a comparison of the worst-case spectral efficiency (SE) of different schemes vs. SNR. The worst-case SE is $ N_{\rm s} \times {\rm log}_2(1/\rho)$} \cite{palomar2007mimo}. As we can see, our schemes are more efficient than the SDP-based scheme and close to  \cite{wang2022joint} with perfect CSI. Moreover, the FCS is more efficient than the PCS, due to the higher effective channel gain of the former.

\begin{figure}[t]
	\centering
	\includegraphics[width=7 cm,height=5.5 cm]{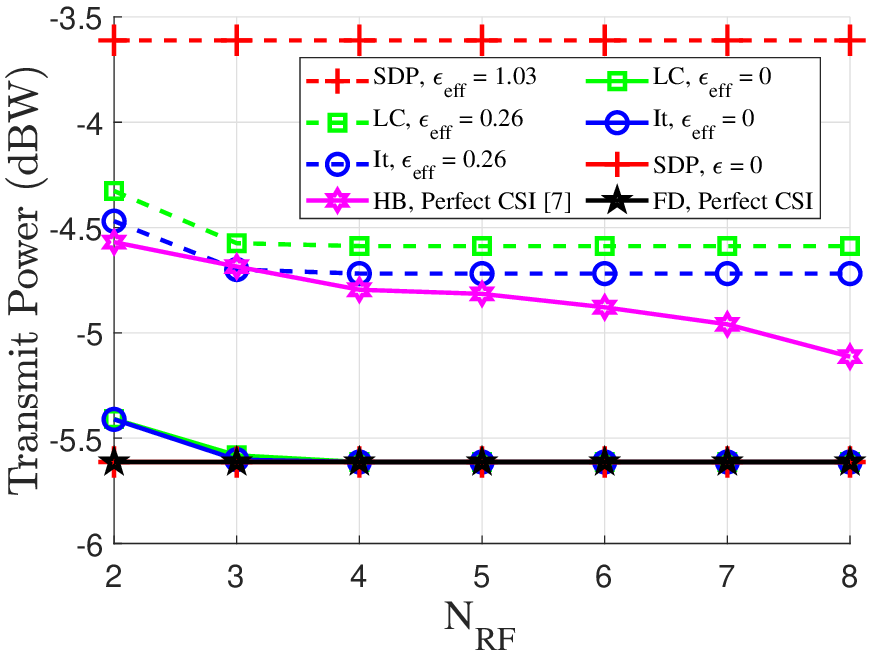}
	\caption{{Transmit power of our HB schemes vs. $N_{\rm RF}$ for $N_{\rm s}=2$.}}
	\label{power_nrf}
\end{figure}

\begin{figure}[!t]
	\centering
	\includegraphics[width=7 cm,height=5.5 cm]{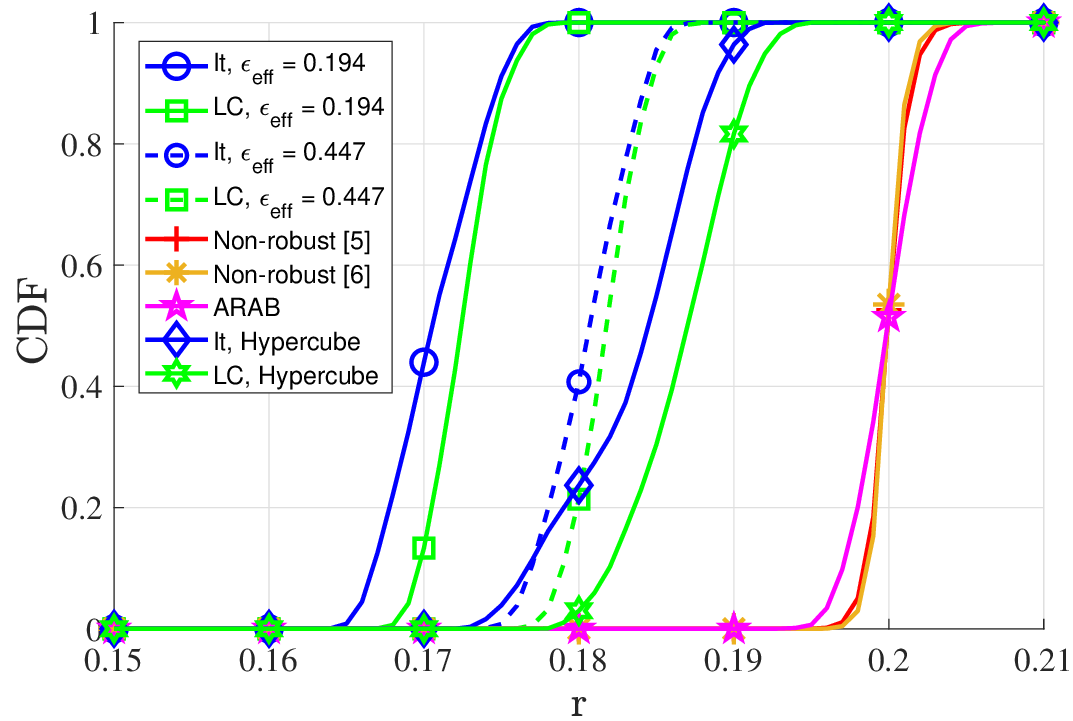}
	\caption{{CDF in our iterative and low complexity HB schemes and in non-robust HB schemes vs. $r$ in ${\rm Pr}\{{\rm MSE}_k \le r \} = 1$.}}
	\label{cdf_mu}
\end{figure}

{As discussed above, we use approximation of channel eigenvectors in the RF precoder and combiner,} where higher values of $N_{\rm{RF}}$ results in more accurate approximations of eigenvectors \cite{tajalli2021qos}. Fig. \ref{power_nrf} shows that our schemes consume less transmit power for higher values of $N_{\rm RF}$. {In other words, the transmit power is less when the eigenvectors of the channel are better approximated. When $N_{\rm RF}\ge 2N_{\rm s}$, with perfect CSI, our schemes consume the same power as the fully digital SDP-based scheme, but with imperfect CSI, our schemes consume less power.}

We now compare the performance of our schemes with that of the non-robust schemes in multi-user systems. Fig. \ref{cdf_mu} shows the CDF of MSEs in our robust schemes and in the non-robust schemes detailed in \cite{ni2015hybrid}, \cite{wu2018hybrid}, and \cite{ardah2020hybrid} (called ARAB). We consider a $64 \times 16$ multi-user system with two users. The base station and the users have $N_{\rm t}^{\rm RF} =4$ and $N_{\rm r}^{\rm RF} =2$ RF chains, respectively. The base station transmits four data streams, i.e., $N_{\rm s} = 2$ data streams intended for each user, with the highest acceptable MSE $\rho = 0.2$. The non-robust HB scheme in \cite{ni2015hybrid} {selects the RF precoder and combiner from the columns of discrete Fourier transform (DFT) matrix. The RF precoder and combiner in the single-stage method} \cite{wu2018hybrid} {are the angles of entries in the eigenvectors of the estimated channel matrix. Both of the non-robust schemes adopt BD in the baseband precoder and MMSE combining. The size of the uncertainty region is $\varepsilon_{\rm eff} = 0.194, 0.447$ corresponding to $P_{\rm in} = 1 - 10^{-3}$ when $\sigma_{\rm e}^2= 0.01, 0.05$.} Fig. \ref{cdf_mu} shows that our schemes satisfy MSE constraints, as ${\rm Pr}\{{\rm MSE}_k \le r \} = 1$ when $r= 0.2$; however, the non-robust schemes fail when the error matrices are in the uncertainty region of size $\varepsilon = 3.357$, which is equivalent to $\varepsilon_{\rm eff} = 0.447$. More specifically, the CDF is more than $0.5$ at $r =0.2$; that is, the non-robust schemes fail to satisfy MSE constraints in about 50\% of channel realizations. Fig. \ref{cdf_mu} also shows the robustness of our schemes for error models that are not norm-bounded. For instance, our schemes are able to satisfy MSE constraints when channel error is in a hypercube uncertainty region defined by $\mathcal{R}\{[\Delta]_{i,j}\}\le \frac{\varepsilon}{\sqrt{2N_tN_r}}$ and $\mathcal{I}\{[\Delta]_{i,j}\}\le \frac{\varepsilon}{\sqrt{2N_tN_r}}$.

\begin{figure}
	\centering
	\includegraphics[width=7 cm,height=5.5 cm]{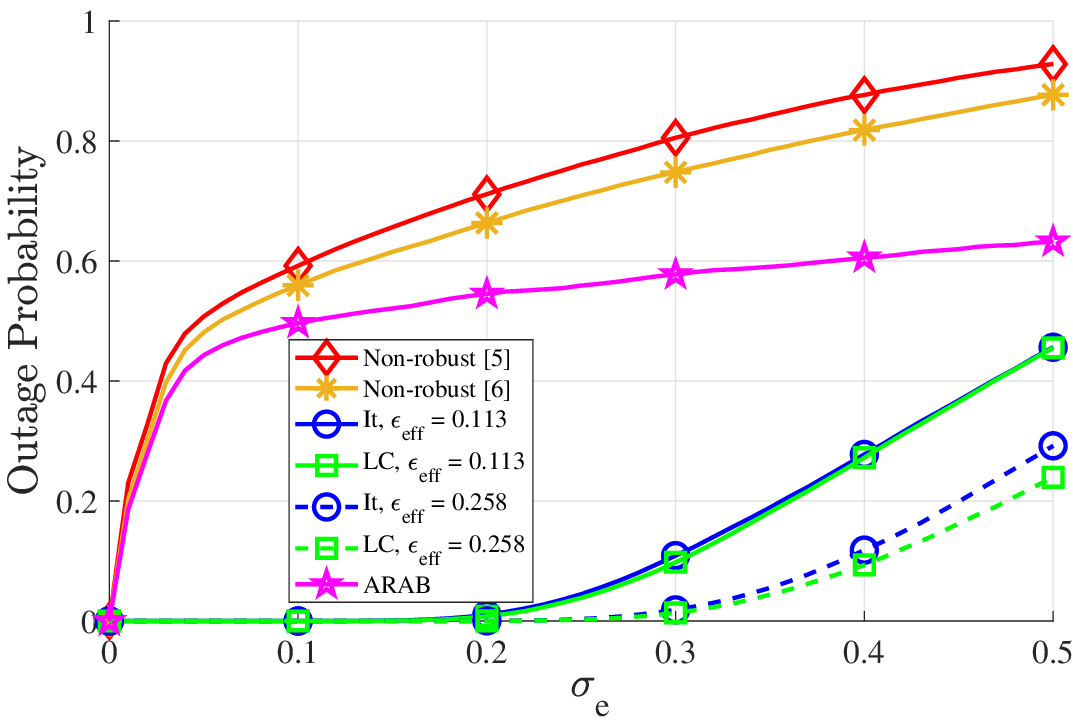}
	\caption{{Outage probability in our robust HB scheme and in the state-of-the-art non-robust schemes.}}
	\label{outage_mu}
	\end{figure}

\begin{figure}[!t]
	\centering
	\includegraphics[width=7 cm,height=5.5 cm]{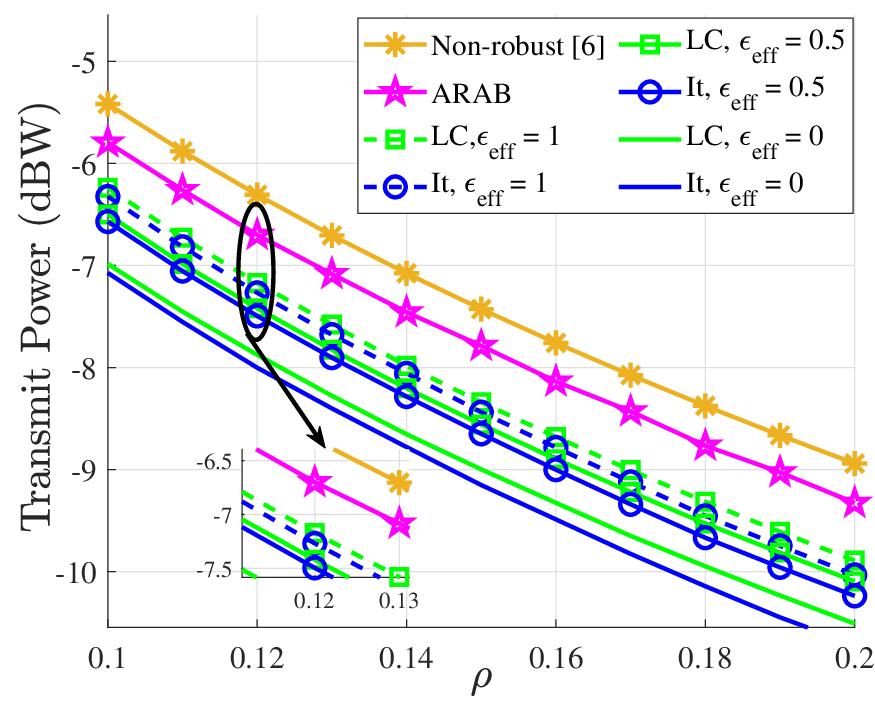}
	\caption{{Transmit power in our iterative and low complexity HB schemes and in non-robust HB schemes.}}
	\label{txpower_mu}
\end{figure}

Fig. \ref{outage_mu} {shows the probability of an outage when error matrices are not necessarily in the uncertainty region. The probability of an outage ${\rm Pr}\{{\rm MSE}_k > 0.2 \}$ in the non-robust schemes significantly increases as $\sigma_{\rm e}$ becomes larger. By contrast, the probability of outage in our schemes is zero for small values of $\sigma_{\rm e}$, smoothly increases for higher values of $\sigma_{\rm e}$, and is smaller for larger uncertainty regions.}

Fig. \ref{txpower_mu} {shows the transmit power in our robust schemes and in the state-of-the-art non-robust schemes versus the highest acceptable MSE $\rho$ in a $100 \times 16$ multi-user system with four users.} The base station and each user have $N_{\rm t}^{\rm RF} = 12$ and $N_{\rm r}^{\rm RF} = 3$ RF chains, respectively. The base station transmits $N_{\rm s} = 2$ data streams to each user. When $\varepsilon_{\rm eff} = 0$, our schemes consume less transmit power than \cite{ardah2020hybrid} (ARAB) and the non-robust scheme in \cite{wu2018hybrid}. Notably, our robust schemes outperform \cite{wu2018hybrid} and ARAB, even in relatively large uncertainty regions $\varepsilon_{\rm eff} = 0.5, 1$. In addition, the transmit power is higher for larger values of $\varepsilon_{\rm eff}$, which is due to a smaller feasible region in \eqref{qos_bb_problem} for larger $\varepsilon_{\rm eff}$.

\begin{figure}[!t]
	\centering
	\includegraphics[width=7 cm,height=5.5 cm]{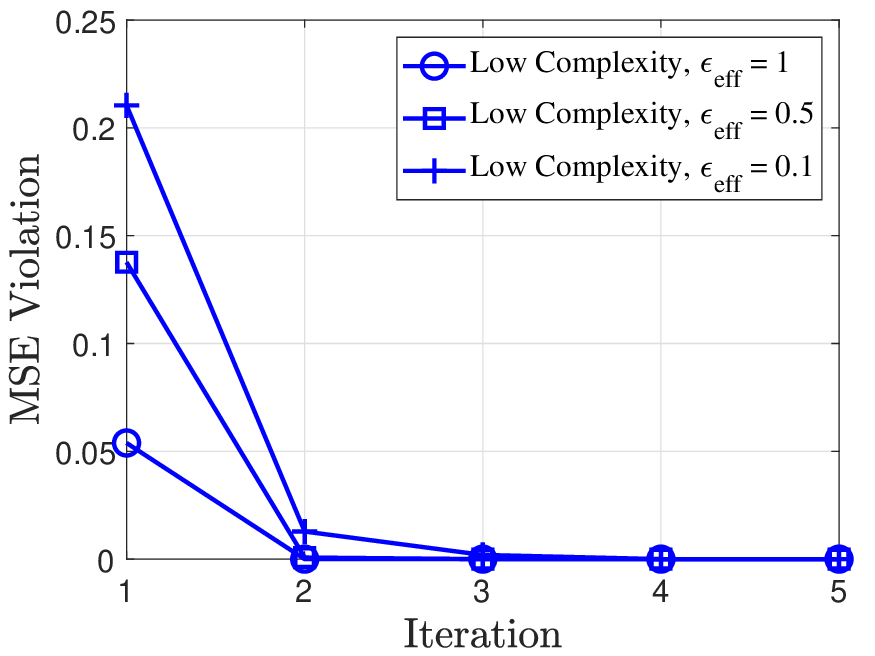}
	\caption{MSE violation of the cutting-set method for different values of uncertainty in the effective channel $\varepsilon_{\rm eff}$.}
	\label{convergence}
\end{figure}

Fig. \ref{convergence} shows the convergence of the cutting-set method for a $128 \times 36$ multi-user system with three users. The base station transmits $N_{\rm s}=3$ data streams for each user. The base station and each user are equipped with $N_{\rm t}^{\rm RF} = 9$ and $N_{\rm r}^{\rm RF} = 3$ RF chains, respectively. The convergence is analyzed by computing MSE violations for each data stream in different iterations of the cutting-set method. {An MSE violation} is defined as $V = \sum_{u=1}^U\sum_{k=1}^{N_{\rm s}}({\rm MSE}_{u,k}^{\rm wc}-\rho_{u,k})$. Fig. \ref{convergence} shows that the cutting-set method converges in only two or three iterations.

\section{Conclusion}\label{conclusion_section}

In this paper, we developed two robust QoS-aware HB schemes for mmWave massive MIMO systems. {More specifically, we proposed an iterative scheme, based on the cutting-set method that includes alternating power minimization and worst-case analysis. We also proposed a low-complexity scheme that diagonilizes the estimated channel in the baseband precoder, and used MMSE combining in the baseband combiner; and we obtained closed-form solutions for the transmit power. As we saw, our schemes outperform the fully digital robust beamforming and conventional non-robust HB. Our low-complexity scheme was also shown to have a higher acceptance ratio with more transmit power than our iterative scheme, although less than the other schemes. We also showed rapid convergence of the cutting-set method.}

\ifCLASSOPTIONcaptionsoff
  \newpage
\fi

\end{document}